\documentclass[preprint]{aastex}

\def\etal   {{et~al.\/}}
\def\kms    {~km~s$^{-1}$}

\begin{document}
\input psfig.sty

\title{Near-Infrared Spectroscopy of Two Galaxies at $z=2.3$ and $z=2.9$:
New Probes of Chemical and Dynamical Evolution at High Redshift\altaffilmark{1}}

\author{Henry A. Kobulnicky\altaffilmark{2}}
\affil{Astronomy Department, University of Wisconsin - Madison}
\affil{475 N. Charter St., Madison, WI 53706}
\altaffiltext{1}{Based on observations obtained at the W.~M. Keck Observatory 
which is operated jointly by the University of California and the 
California Institute of Technology.}
\altaffiltext{2}{Hubble Fellow}
\email{chip@astro.wisc.edu}
\author{David Koo}
\affil{UCO/Lick Observatory}
\affil{Department of Astronomy and Astrophysics}
\affil{University of California, Santa Cruz}
\affil{Santa Cruz, CA 95064}
\email{koo@ucolick.org}

\author{Final Draft of 2000 August 15}

\begin{abstract}

This study presents Keck optical and infrared spectroscopy of the
rest-frame ultraviolet and optical emission lines in two Lyman $\alpha$
emitting galaxies at $z>2$.  These data provide insight on the
evolution of fundamental galaxy scaling relations at early epochs,
especially the luminosity-velocity and luminosity-metallicity
relations.  Spectral diagnostics suggest that the Coup Fourr\'e Galaxy
at $z=2.3$ [CFg; Lowenthal \etal\ 1991] and Lynx~2-9691, a
serendipitously-discovered, luminous Lyman-drop galaxy at $z=2.9$, are
star-forming galaxies without active nuclei.  Lynx~2-9691 exhibits
extended [O~III] emission over a diameter of $>$28 kpc, reminiscent of
the Lyman $\alpha$ nebulae discovered near Lyman-drop galaxies [Steidel
\etal\ 2000].  We estimate star formation rates of 59 $M_\odot~yr^{-1}$
and 111 $M_\odot~yr^{-1}$, respectively, from Balmer recombination line
luminosities, 2-3 times higher than inferred from the ultraviolet
continuum.  The ratios of strong nebular emission lines indicate
sub-solar oxygen abundances in the range $8.2<12+log(O/H)<8.8$ ($Z =
0.25-0.95~Z_\odot$).  Interestingly, Galactic metal-rich globular
clusters have similar metallicities, consistent with the idea that we
could be seeing the formation of galaxies like the Milky Way at
$z\sim3$.  The measured gas phase oxygen abundances are $>4-10$ times
higher than the $Z < 0.1 Z_\odot$ metallicities found in damped Lyman
$\alpha$ (DLA) absorbers at similar redshifts, indicating that DLA systems trace
fundamentally different environments than the vigorously star-forming
objects observed here.  If this intense star formation activity
represents the dominant formation episodes for stars in today's spiral
bulges or ellipticals, then the evolved descendants in the local
universe should exhibit similarly sub-solar metallicities in their
dominant stellar populations which formed 8-10 Gyr ago.  When these new
data are combined with a sample of four other high-redshift spectroscopic
results from the literature, we find that star-forming galaxies at
$z\sim3$ are 2-4 magnitudes more luminous than local spiral galaxies of
similar metallicity, and thus, are offset from the local
luminosity-metallicity relation.  Their kinematic linewidths are
$\sigma_v=65-130$ \kms, making this sample 1-3 magnitudes more
luminous than local galaxies of similar linewidth and mass.  Less
luminous Lyman-drop galaxies need to be studied to see if these
deviations are universal or apply to only the most luminous
high-redshift galaxies.

\end{abstract}
\subjectheadings{galaxies: formation --- galaxies: evolution ---
galaxies: fundamental parameters --- galaxies: abundances --- Galaxy: abundances
--- Galaxy: globular clusters }

\section{Introduction}

The early epochs of galaxy formation have recently become accessible to
direct observation (see reviews by Dickinson 2000; Steidel 1999; Stern
\& Spinrad 1999).  Of particular importance are the so-called ``Lyman
drop'' galaxies\footnote{We prefer to use the term ``Lyman-drop''
instead of ``Lyman break''  since the intergalactic Lyman $\alpha$
forest absorption, and not just the Lyman break at 912 \AA\ may also
contribute to a decrease in the far UV flux, especially at higher
redshifts.  See Lowenthal \etal\ 1997 where this is demonstrated
empirically by the use of B-band in addition to U-band dropouts to
select $z\sim2.5$ galaxies.} at redshifts $z>2$ with high star
formation rates suggesting that they may be the progenitors (Steidel
\etal\ 1996a,b) or building blocks (Lowenthal 1997) of today's
ellipticals, spiral bulges, and spiral halos.  Virial masses determined
from the optical emission linewidths have lower limits of a
few$\times10^{10}~M_\odot$.  Balmer line luminosities indicate star
formation rates (SFR) of 20-270 $M_\odot~yr^{-1}$, a factor of several
times higher than the SFR implied from their ultraviolet continuum
luminosities (Pettini \etal\ 1998).   During this period 8-10 Gyr ago,
it appears that many of today's large cluster galaxies and the universe
as a whole (Madau \etal\ 1996, Steidel \etal\ 1999) were undergoing
their dominant episodes of star formation.

Infrared spectroscopy in the near-IR windows provides a view of the
rest-frame optical properties of galaxies in this active redshift range,
$2.0<z<3.5$.  One application has been the determination of gas-phase
metallicities within galaxies over the same redshift ranges where
damped Lyman $\alpha$ (DLA) systems show metallicities
$\leq0.1~Z_\odot$ (Pettini \etal\ 1997a,b;1999; Lu \etal\ 1996).  DLA 
systems show little or no change in metallicity from
$z=3$ to the present (Prochaska \& Wolf 1999, 2000; Pettini
\etal\ 1999), in conflict with theoretical expectations (Pei, Fall, \&
Hauser 1999, although see Cen \& Ostriker 1999).   Another application
is to understand how fundamental galaxy parameters like size, mass,
luminosity, and metallicity scale when the universe was just 10\% its
present age and galaxies were in their youth.

These types of studies are only beginning with the advent of near-IR
spectrographs on 8-10 m class telescopes.  Teplitz (2000a,b) presented
infrared spectra of strong nebular lines in two $z>3$ Lyman-drop
galaxies.  They derived slightly sub-solar gas-phase metallicities in
the range 0.2-0.9 $Z_\odot$ and star formation rates several times
higher than those inferred from the ultraviolet continuum
luminosities.  In this paper we report optical and infrared
spectroscopy for two additional objects at $z>2.3$: Lynx~2-9691, a
serendipitously-discovered luminous Lyman-drop galaxy and another
Lyman $\alpha$ emission galaxy known as the ``Coup Fourr\'e Galaxy''
(CFg; Lowenthal \etal\ 1991; Roche, Lowenthal, \& Woodgate 2000).  In
all luminosity-related computations, we assume a cosmology with
$H_0=70~km~s^{-1}~Mpc^{-1}$, $\Omega_0=0.2$, and
$\Omega_\Lambda=0$.\footnote{Adopting instead a flat 
$\Omega_\Lambda=0.7$ cosmology
would change the derived luminosities by $<$0.2 mag. at $z=3$.}

\section{Data Acquisition}

The two high-redshift objects were observed in the optical and infrared
spectral regions at the Keck II telescope.  Using the NIRSPEC infrared
spectrograph (McLean \etal\ 2000) during UT 2000 January 23--24, we
obtained spectra of Lynx~2-9691 and the CFg.  Table~1 lists positions,
redshifts, and apparent magnitudes for the two target objects. 

The CFg was discovered as a strong Lyman $\alpha$ emitter
48\arcsec\ from QSO PHL 957, which possesses a damped Lyman $\alpha$
system at the same redshift.  The CFg has a redshift of $z=2.3128$
(Lowenthal \etal\ 1991) and is composed of two components, CFgA and
CFgB, separated by 0.35\arcsec.  For our purposes, we consider both
components together as a single galaxy.  Lynx~2-9691 was discovered
serendipitously during a Keck study of the chemical compositions of
$z\sim0.4$ galaxies in the Lynx~2 field (Kobulnicky \& Zaritsky 1999).
During an optical observing run with the LRIS spectrograph (Oke 1995)
on Keck II during 1997 November, we obtained the discovery spectrum of
Lynx~2-9691 over the wavelength range 3650 \AA -- 8000 \AA.  Details of
the observations and reductions are given in Kobulnicky \& Zaritsky
(1999).  The spectrum has a resolution of 9 \AA\ FWHM, an integration
time of 1.5 hours, and is flux calibrated based on spectrophotometric
standards.  Lynx~2-9691 appears as a star-like object near the limit of
$R_F\simeq23$ on deep 4-m photographic plates and lies at $z=2.8877$.
Figure~1a shows a 37\arcsec\ image of the field from scans of
photographic plates (M. Bershady, private communication) taken with the
$R_F$ filter (Kron 1980).  Rectangles mark the positions of Keck LRIS
slitlet used in the discovery observations and the Keck/NIRSPEC slit
used for IR spectroscopy.  Figure~1b shows the 2-D Keck/LRIS discovery
spectrum of Lynx~2-9691.  Strong Lyman $\alpha$ emission is seen near
$z=2.9$.  With an observed  magnitude of $R_F\sim23$ (Table~1), it is
among the most luminous Lyman-drop galaxies when compared to the sample
of Steidel \etal\ (1996, 1998).  Lynx~2-9691, however, may be lensed,
since it lies within a few arcsec of several other faint galaxies,
including an early type galaxy, Lynx~2-9689, located 3.5\arcsec\ to the
NE, which shows strong stellar Ca H\&K absorption features at
$z=0.31$.  High-quality imaging of the field and measured redshifts for
the other field galaxies would be useful to address this possibility.
Unless otherwise noted, we assume that there is no magnification due to
lensing.   In any case, lensing will not change the measured chemical
composition or kinematic linewidth, only the apparent luminosity and
size of the system.

We obtained low resolution $R\simeq1300$ spectra of each object in
three near-infrared passbands selected to contain the redshifted
emission lines of [O~II] $\lambda$3727, H$\beta$, [O~III]
$\lambda$5007, and H$\alpha$.  In the case of the CFg, these passbands
correspond to the standard {\it JHK} filters.  For Lynx~2-9691, the NIRSPEC
N4, N6, and N7 filters were used.  The position angle of the slit was
46 degrees for the CFg observations and 118 degrees for Lynx~2-9691.
Table 2 summarizes the wavelength coverage, integration times, and
other relevant observational details.  Between two and five
exposures measuring 300 s to 900 s were obtained of each object.
Airmasses ranged between 1.11 and 1.55.  In all cases, the 42\arcsec x
0.76\arcsec\ longslit, the widest available, was used.  The scale 
of the 1024x1024 detector is
0.178\arcsec/pix in the spatial direction and 0.14\arcsec/pix in
the spectral direction.  Between exposures, the telescope was dithered
(i.e., repositioned) to move the object several arcseconds along the
slit.  Seeing varied between 0.6\arcsec and 1.1\arcsec.  Both nights
were clear, but windy.

\subsection{Reductions and Calibration}

The NIRSPEC data reduction in IRAF\footnote{IRAF is distributed by the National
Optical Astronomy Observatories, which are operated by the Association
of Universities for Research in Astronomy, Inc., under cooperative
agreement with the National Science Foundation.} involved division by a
halogen lamp flat field to remove pixel sensitivity variations in the
spectral direction.  To remove the strong night sky emission lines and
background, we subtracted from each frame the dithered frame
immediately preceding or following each exposure.  Wavelength
calibration was performed using the OH airglow lines present in
each frame.  The wavelength solution was good to an RMS of 2 \AA.  The
output pixel scale was $\sim3$ \AA/pix.  Gaussian fits to individual
night sky lines indicate a mean spectral resolution of 13 \AA\ FWHM at
1.6 $\mu$m or $R\sim1250$.  Observations of the twilight sky were used
to correct the spectra for response variations along the slit.  In
practice, all objects were observed within 10\arcsec\ of the slit
center where the response is uniform.

Spectra were aligned spatially by shifting an integer number of pixels
along the slit.  In most cases, offsets were determined directly from
emission lines visible in the individual background-subtracted frames.
No continuum from the target galaxies was detected.  When no emission
lines were present, offsets were determined from stars in the slit
viewing camera (SCAM) images.  Similar images were combined by
averaging with rejection for cosmic rays.  Spectral extraction was
accomplished with the aid of spatial traces from standard star
exposures.

Flux calibration is based upon observations of the spectrophotometric
standard star Feige 34 (Oke 1990) observed at airmass 1.40 and of a
V=10.02 K4V star, HIP 23694 ($\equiv$GSC 03734-00980), observed at
airmass 1.20.  We obtained the absolute flux calibration of Feige 34
using the the STScI STIS/NICMOS optical calibration of hot white dwarfs
(see Bohlin 2000; Bohlin 1996) and modeling Feige 34 as a blackbody of
temperature 67,000 K (Thejll, Ulla, \& MacDonald 1995) to extrapolate
from the optical into the infrared bands.   This approach was adopted
after comparing model blackbody curves to the optical and infrared
observations of another similarly hot white dwarf, G191B2b.  This
approach, however, neglects the small amplitude Bracket absorption
lines in the atmosphere of the white dwarf.  Fortunately, these
absorption lines do not fall at the wavelengths where we are
calibrating emission lines in the high-redshift targets, so there is no
impact on the flux calibration.  For HIP 23694, we use the spectral
energy distribution of a K4V star (Pickles 1988) and normalize the flux
using V=10.02 (Hipparchos Catalog).  Thus, we have two independent flux
calibrations which agreed to within 10\% over the range of bandpasses
used here.  For each star, we used the ratio of the model to the
observed spectrum to produce a flux calibration curve which converts
raw count rates into flux calibrated spectra.  This correction factor
implicitly includes the corrections for instrumental response as a
function of wavelength, as well as atmospheric transmission effects
{\it at one particular airmass.} Because atmospheric opacity changes
with airmass, there will be small residuals in the flux calibration for
spectral regions with large opacity.  Since the standard star
observations bracket the range of airmasses for the targets, these
residual factors should be small.

In summary, the {\it relative} flux calibration {\it within} a given
bandpass appears to be good to a few percent in regions uncontaminated
by night sky emission lines or atmospheric absorption features.  The
relative calibration {\it across bandpasses} is accurate to about
15\%.  The zero point of the flux calibration may overestimate the flux
of the targets due to slit losses during observation of the standard
stars.  However, slit losses during target observations will work in
the opposite direction to reduce, or even underestimate, the flux of
sources.   The seeing averaged 0.6-1.1\arcsec\ while the maximum slit
width for NIRSPEC in longslit mode is 0.76\arcsec.  Since we are
primarily interested in the {\it relative} flux between one spectral
line and another, the calibration is adequate for the present
purposes.

\section{Results and Analysis}

The extracted spectra of the targets appear in Figures 2--8.  Figure 2
is the optical (rest-frame ultraviolet) spectrum of CFg, from Lowenthal
\etal\ (1991).  The prominent Lyman $\alpha$ line is marked along with
He~II~$\lambda$1640 and C~IV~$\lambda$1549.  A sample night sky
spectrum appears below, scaled arbitrarily, for reference purposes.
Lowenthal \etal\ (1991) argue that this is a normal star-forming
galaxy, rather than an AGN, on the basis of the small velocity width
($<$700 \kms) and lack of high ionization lines like N~V $\lambda1240$
normally seen in Seyfert galaxies.  High resolution HST imaging shows
no evidence of a central point source that would indicate nuclear
activity (Roche \etal\ 2000).  However, the presence of He II $\lambda$
1640 and C~IV $\lambda$1549 in emission suggest a high degree of
ionization seen only in the most energetic star formation regions.

Figure 3 shows the optical (rest-frame ultraviolet) spectrum of
Lynx~2-9691 from rest wavelengths 950 \AA--2050 \AA.  Night sky airglow
lines appear below, scaled arbitrarily.  The expected positions of
common nebular emission and absorption lines are marked.  Most
prominent is Lyman $\alpha$ with a rest-frame equivalent width of 18
\AA.  The reported equivalent widths are corrected to the rest frame
using

\begin{equation}
EW_{rest} = EW_{observed}/(1+z) .
\end{equation}
 
\noindent  He II
$\lambda$1640 is present in emission, and possibly
C~III] $\lambda$1909.  If so, both 
are indicators of a high ionization level, probably
due to massive stars.  The most prominent absorption features are C~IV
$\lambda$1548/51 and Si IV $\lambda$1394/1403.  These may have either
interstellar or stellar origins.

Figure~4 shows an expanded portion of the Lynx~2 spectrum in the far
ultraviolet.  Nearly all of the observed features are common in
atmospheres of O and Wolf-Rayet stars seen with IUE  (Neubig \& Bruhweiler
1999) and the Hopkins Ultraviolet Telescope (Schulte-Ladbeck, Hillier,
\& Herald 1995).  The presence of these strong stellar/interstellar
features, and the absence of broad, high-ionization features is
consistent with normal star-forming galaxies containing massive
stars.  The spectra in Figures 3 and 4 bear many similarities to
the Lyman-drop galaxies, argued to be normal star-forming galaxies by
Steidel \etal\ (1996a,b).

Figure~5 shows the sky-subtracted 2-D spectra for the CFg.  Positions of
the strong nebular lines are labeled.  Figure~6 shows the extracted 1-D
infrared J, H, K band spectra of CFg.  Night sky emission lines appear
below each target spectrum, scaled arbitrarily.  An atmospheric
transmissivity curve based on models for Mauna Kea (Lord 1992) is
plotted above each spectrum, scaled arbitrarily, with a horizontal
scale bar indicating zero transmissivity.  These curves serve to show
that the major emission lines are not strongly affected by regions of
atmospheric absorption.  Marks denote the expected redshifted positions
of [O~II] $\lambda$3727, H$\beta$, [O~III] $\lambda\lambda$4959/5007,
and H$\alpha$.  [O~III] $\lambda$5007 and H$\alpha$ are clearly
detected, while [O~III] $\lambda$4959 is a less certain detection due
to contamination by neighboring OH airglow lines.  [N~II] $\lambda$6584
is not detected as it falls at the position of a strong night sky line
just redward of $H\alpha$.  H$\beta$ is not detected, even though it
falls in between strong night sky lines.  $H\beta$ is susceptible to
absorption by the underlying stellar population, an effect which may
contribute to its non-detection.  [O~II] $\lambda$3727 lies in a region
contaminated by sky lines, and is not unambiguously detected.  No
continuum is present in any of the infrared spectra.  Extrapolating the
Lowenthal \etal\ (1991) spectral energy distribution from the
ultraviolet to the optical indicates that the continuum flux would be
$<1.0x10^{-18}~erg~s^{-1}~cm^{-2}$\AA$^{-1}$, below the 3$\sigma$
detection limit $<6.0x10^{-18}~erg~s^{-1}~cm^{-2}$\AA$^{-1}$ in the H
band (neglecting dust extinction).

Figure~7 shows the sky-subtracted 2-D spectra for Lynx 2-9619.  Labels
mark the positions of the strong nebular lines.  Figure~8 shows the
infrared NIRSPEC N4, N6, N7 band spectra of Lynx~2-9691.  Night sky
emission lines appear below each target spectrum, scaled arbitrarily.
An atmospheric transmissivity curve based on models for Mauna Kea (Lord
1992) is plotted above each spectrum, scaled arbitrarily, with a
horizontal scale bar indicating zero transmissivity.  Marks denote the
expected redshifted positions of [O~II] $\lambda$3727, H$\beta$,
[O~III] $\lambda\lambda$4959/5007, and H$\alpha$.  In some cases, poor
atmospheric transmissivity strongly affects the detectability of the
line.  H$\alpha$ is lost amidst the night sky lines and strong
atmospheric absorption longward of 2.5 $\mu$m.  [O~III] $\lambda$5007
is a strong detection, while [O~III] $\lambda$4959 is not detected.  A
night sky line near the position of H$\beta$ makes measurement of this
line difficult, though emission is clearly seen on the 2-D image in
Figure~7.  The spike near [O~II] $\lambda$3727 appears to be a residual
cosmic ray, as no emission lines are present in the two-dimensional
spectra.  As with CFg, no continuum is present.  Extrapolation from the
UV continuum to the optical implies a flux level
$<1.0x10^{-18}~erg~s^{-1}~cm^{-2}$\AA$^{-1}$, below the 3$\sigma$
detection limit $<7.0x10^{-18}~erg~s^{-1}~cm^{-2}$\AA$^{-1}$ in the N6
filter, well below the detection limit.  Interestingly,  [O~III]
$\lambda$5007 line is spatially extended over 20 pixels (3.6\arcsec=28
kpc) with a Gaussian FWHM of 2.0\arcsec (16 kpc).  This extended size
is consistent on all three 900~s exposures.  Images on the slit viewing
camera show no signs of tracking problems, so we believe that this
extension is real.

We measured the fluxes and velocity widths of emission lines using
single 1-D Gaussian fits.  Table~3 lists the values and upper limits,
where appropriate.  Since no continuum is detected in the infrared, the
equivalent widths are undefined.   From the $H\alpha$ fluxes, we
compute the $H\alpha$ luminosities and star formation rates using the
calibration of Kennicutt (1983).   Table~3 summarizes these derived
physical parameters for each system.  For the CFg, the value of
$F_{H\alpha}=1.65\pm0.1~\times10^{-16}~erg~s^{-1}~cm^{-2}$ is in
agreement with the narrowband imaging result of Bunker
\etal\ (1995), who find
$F_{H\alpha}=2.1\pm0.6~\times10^{-16}~erg~s^{-1}~cm^{-2}$.  For our
adopted cosmology, the star formation rates are 59 $M_\odot~yr^{-1}$ and
111 $M_\odot~yr^{-1}$ for the CFg and Lynx~2-9691, respectively.  These
values are 3-4 times higher than the mean value of 16 $M_\odot~yr^{-1}$
for similarly luminous Lyman-drop galaxies (Steidel 1996b), but
consistent with the 20-270 $M_\odot~yr^{-1}$ found by Pettini
\etal\ (1998) from Balmer recombination lines.   In Table~3 we list the
1500 \AA\ specific fluxes and luminosities, and the inferred SFR from
the prescription of Madau, Pozzetti, \& Dickinson (1998),

\begin{equation}
L_{1500}=1.06\times10^{40} \times {{SFR}\over{(M_\odot~yr^{-1})}}
erg~s^{-1}~\rm\AA^{-1}.
\end{equation}

The SFRs derived from the
Balmer lines in the CFg and Lynx~2-9691 are factors of $\sim2-3$ times
higher than those derived from the ultraviolet continuum specific
luminosity at 1500 \AA, 31 $M_\odot~yr^{-1}$ for the CFg and 41 $M_\odot~yr^{-1}$
for Lynx~2-9691.  These differences may either imply significant
amounts of dust extinction, or a systematic problem relating
 ultraviolet continuum fluxes to instantaneous star formation rates
(see Glazebrook \etal\ 1998).

We calculate the intrinsic velocity width of the strongest emission
line after the instrumental profile, defined by the width of night sky
lines, has been subtracted in quadrature.  This procedure makes the
simplifying assumption that, due to seeing and target structure, the
galaxy emission line is distributed uniformly across the width of the
slit.  For the CFg, the H$\alpha$ line shows a FWHM of 301 \kms.  The FWHM
of [O~III] is in good agreement at 250 \kms.  Lynx~2-9691 has FWHM=156
\kms\ measured from the [O~III]  $\lambda$5007 line.  Using the
calculated velocity width, we estimate a {\it lower limit} for the
virial mass of each system.  As a fiducial dynamical size, we adopt the
half-light radius (CFg) or the extent of the emission line gas
(Lynx~2-9691).  To estimate virial masses, we use the velocity full
width at 20\% of peak, $W_{20}=3.62\sigma$ and half-light radius,
$r_{hl}$, from Table~1 assuming $M_{vir}=
{{(0.5W_{20})^2~r_{hl}}\over{G}}$ (e.g., Longair 1998), where $G$ is
the gravitational constant.  The derived  dynamical masses are
$M_{vir}\geq7.0\times10^{10}~M_\odot$ for the CFg and
$M_{vir}\geq5.2\times10^{10}~M_\odot$ for L2-9691.  We stress that
these dynamical masses are highly uncertain, in part because we lack
information on inclination which make these values lower limits.
Moreover, the dynamical size scale is difficult to constrain, the
dynamical state of the system may not be in equilibrium, and the
kinematic structure of the objects remains unknown.

\section{Chemical Analysis}

We have calculated an oxygen abundance for each object using
the measured emission line strengths.  Kobulnicky, Kennicutt, \& Pizagno
(1999) showed that whole-galaxy emission line ratios can provide a
reliable indication of the system's nebular oxygen abundance using the
empirical line flux ratio technique of Pagel \etal\ (1979).

\begin{equation}
{R_{23}}\equiv\ \Biggr({{[O~II] \lambda3727 + [O~III]
\lambda\lambda4959,5007}\over{H\beta}}\Biggr).
\end{equation}

Several authors have published
calibrations relating $R_{23}$ to the oxygen abundance, O/H, including
McGaugh (1991), Zaritsky, Kennicutt, \& Huchra, (1994), 
McCall, Rybski, \& Shields (1985), Edmunds
\& Pagel (1984), and Dopita \& Evans (1986).  Here, we adopt 
the McGaugh (1991) calibration, which is similar to the others
over most metallicity regimes (see comparison Figure in
Kobulnicky \etal\ 1999), and it has an analytic approximation making
abundance computation more straightforward.
From the line fluxes listed in Table~3, we have computed $R_{23}$
along with the ionization parameter indicator $O_{32}$:

\begin{equation}
\log(O_{32}) \equiv\ \log\Biggr(
	{{[O~III]\lambda\lambda4959,5007}\over{[O~II]\lambda3727}}\Biggr).
\end{equation}

\noindent Table~4 lists these line diagnostics for each object.  In
addition to the two objects observed here, we include 5 other infrared
spectroscopic results at $z>2$ from the literature (Teplitz
\etal\ 2000a,b; Pettini \etal\ 1998).  Where [O~II] $\lambda$3727 is
not detected, we adopt $F_{3727}=0.0$ and list a lower limit for
$R_{23}$.  When [O~III] $\lambda$4959 is not detected, we assume the
theoretical ratio $F_{4959}=F_{5007}/2.9$.  For the CFg, where H$\alpha$ is
detected but $H\beta$ is not, we adopt $I_{H\beta}=I_{H\alpha}/2.86$
assuming theoretical case B hydrogen recombination ratios for gas with
an electron temperature of 10,000 K (Hummer \& Story 1987).  This
approximation is very insensitive to the actual electron temperature,
but ignores effects of dust extinction.  Any extinction correction
would serve to raise $R_{23}$ by boosting the strength of [O~II]
$\lambda$3727 relative to H$\beta$.  We have also neglected the effects
of stellar absorption on the H$\beta$ line.  Since no continuum is
detected, we cannot even apply a statistical correction of 2--4 \AA\ in
the equivalent width, which is representative of local spiral galaxies
(Kennicutt 1992).  Thus, without many of the traditional constraints,
any metallicity determination carries a significant uncertainty.  We
proceed to evaluate the results, while advocating a prudent amount of
caution.

Figure~9 shows the diagnostic diagram $R_{23}$ versus oxygen abundance,
12+log(O/H).  The solid lines are models from McGaugh (1991) for
realistic ranges of ionization parameter represented by
$\log(O_{32})$.  Each object from Table~4 is plotted either as an
$R_{23}$ lower limit, or a shaded box representing the probable
location.  The CFg is plotted as a lower limit with $\log{R_{23}}>0.86$.
For this object, we measure [O~III] $\lambda$5007 and $H\alpha$.  We
infer $H\beta$ from H$\alpha$.  There is no constraint on the
extinction, and thus the actual $H\beta$ flux could be substantially
less than the value adopted from the theoretical $H\alpha$/$H\beta$
ratio.  The $R_{23}$ lower limit constrains the oxygen abundance of the
CFg to be $7.8<12+log(O/H)<8.7$.  In computing the $R_{23}$ lower
limit, we have adopted $F_{3727}=0$, while the measured upper limit is
$F_{3727}<0.47\times10^{-16}~erg~s^{-1}~cm^{-2}$.
For Lynx~2-9691, we have detected [O~III] $\lambda$5007 at a high level
of significance and H$\beta$ at about the $3\sigma$ level.  In
computing the limits on $R_{23}$, we have constrained $F_{3727}$ to lie
between 0.0 and the $3\sigma$ upper limit of
$F_{3727}<0.50\times10^{-16}~erg~s^{-1}~cm^{-2}$.  H$\alpha$ lies in a
region of high atmospheric opacity and is not detected.  With these
values, we constrain $R_{23}$ to lie between $0.70<R_{23}<1.02$ while
$\log(O_{32})>0.77$.  Thus, we plot Lynx~2-9691 as a shaded strip in
Figure~9, with an oxygen abundance in the range $8.3<12+log(O/H)<8.8$.
For the moment, we have assumed that Lynx~2-9691 lies on the upper
(metal-rich) branch of the empirical calibration.  
The other 4 targets in the literature, summarized in Table~4, have only
lower limits on $R_{23}$, and thus we can only estimate their oxygen
abundances within a broad range (see Figure~9).  The best constrained
is Q0201+113 C6 from Pettini \etal\ (1998) for which
$7.7<12+log(O/H)<8.8$.

In principle, the metallicity of Lynx~2-9691 could also lie along
conjugate position on the lower branch of the calibration, leading to
$7.7<12+log(O/H)<8.0$.  We argue that such extremely low metallicities
are improbable for objects like Lynx~2-9691 and the luminous Lyman-drop
galaxy population.  The first piece of evidence is based upon the
metallicity of MS1512-cB58, the first high-redshift object in the
literature with a strong metallicity constraint (Teplitz
\etal\ 2000a).  As a highly-magnified Lyman-drop galaxy, all of the
strong nebular emission lines are observed in MS1512-cB58, yielding
$R_{23}=0.92\pm0.07$ and $\log(O_{32})=0.17\pm0.04$.  The probable
range of the oxygen abundance becomes $8.2<12+log(O/H)<8.5$ consistent
with the CFg and Lynx~2-9691.  The nitrogen-to-oxygen ratio in
MS1512-cB58 is $log(N/O)=-1.24$, indicating significant levels of
chemical enrichment by lower mass stars ($2 M_\odot < M < 6 M_\odot$;
e.g. Pagel 1998; Lattanzio \etal\ 2000).  In the local universe, and
even in the intermediate-redshift universe at $z\sim0.4$ (Kobulnicky \&
Zaritsky 1999), galaxies as metal poor as $7.7<12+log(O/H)<8.0$ also
have low luminosities ($M_B>-18$).  If lower branch metallicities were
adopted for the $z\sim3$ objects, then they would deviate even more
strongly than they do from the local metallicity-linewidth and
metallicity-luminosity relations, as discussed in the next sections.

\section{Discussion of Results}
\subsection{Evolution of Kinematic Properties}

Studying the evolution of fundamental galaxy scaling relations is one
way of understanding how galaxies evolve from the early universe to the
present.  The evolution of the luminosity-linewidth relation (i.e.,
Tully-Fisher) at higher redshifts has been investigated by several
authors (Forbes~\etal\ 1995; Vogt~\etal\ 1996, 1997) who find small
offsets $<$ 0.5 magnitude from the local TF relation out to z=1.0,
while others find more dramatic changes of 1-2 magnitudes (Simard \&
Pritchet 1998; Rix \etal\ 1997; Mallen-Ornelas \etal\ 1999).  In these
studies, galaxies are more luminous for a given linewidth when compared to
local samples.  However, the bluest galaxies deviate most strongly from
the TF relation, suggesting that the conclusions are sensitive to target
selection effects (Bershady~\etal\ 1998).  
With the kinematic linewidths we
derive here, supplemented by the objects from Pettini \etal\ (1998) and
Teplitz \etal\ (2000a) in Table~4, we can begin to assess the status of
the luminosity-velocity correlations at $z>2.3$ and compare them to 
local galaxies.

Bullock \etal\ (2000) parameterize possible forms of the evolution for
the luminosity-velocity correlation as

\begin{equation} M_B = -18.71 - 2.5log\left[
(1+z^\alpha)^{\beta/\alpha}\right]
	- 6.76(log2v_m-2.5) + 5logh \end{equation}

\noindent Here, $z$ is the redshift, $h$ the Hubble constant in units
$H_0/100$, and $v_m$ the maximum circular velocity of the galaxy halo,
which we take to be the velocity on the flat part of the rotation
curve, as in Bullock \etal\ (2000) and Gonzalez \etal\ (2000).  The
parameters $\alpha$ and $\beta$ control the behavior of the relation as
a function of redshift.  In Figure~10 we show the form of the
luminosity-velocity relation for redshifts of $z=0,1,2,$ and 3 when
$\alpha=1$ and $\beta=3$.  We plot the data for a sample of local
galaxies used in Kobulnicky \& Zaritsky (1999; star symbols) and local
luminous HII galaxies (Telles \& Terlevich 1997; open squares).  The
data for the high-redshift galaxies in Table~4 appear as open circles
with error bars.  The local data are corrected for inclination,
but the $z\sim0.4$ and $z\sim3$ samples are only upper limits since 
inclinations are not known.  
While
not intended to be a rigorously-selected sample, these data serve to
illustrate the locus occupied by local galaxies with chemical and
dynamical measurements.  In computing the maximum circular velocities
we adopt the prescription of Tully \& Fouqu\'e (1985),

\begin{equation}
W_R^2 = W_{20}^2 + W_t^2 - 2W_{20}W_t[1-e^{-(W_{20}/W_c)^2}] - 2W_t^2
e^{-(W_{20}/W_c)^2}.
\end{equation}

\noindent Here, $W_R$ is the rotation full amplitude which is 2$\times
V_{max}$.  $W_t=38$ \kms\ is the width due to turbulent motions and
$W_c=120$ \kms\ is the transition point between galaxies having
Gaussian and those having double-horned HI profiles. 

The analytic relations in Figure~10 ($\alpha=1$ and $\beta=3$) are
qualitatively consistent with the data at redshifts beyond
$z=2.3$, in that the zero point of the relation shifts toward 
larger luminosities at higher redshift.  
This figure shows that the high-redshift galaxies lie 1-3
magnitudes more luminous for their velocity width than most normal
local galaxies.  We note that such deviations can be found among
luminous HII galaxies locally (Telles \& Terlevich 1997; open squares)
or among compact galaxies at intermediate redshift (Fig. 3 of Koo
\etal\ 1995).  Although there is some uncertainty as to whether the
integrated emission linewidths reflect the full dynamical width of the
galaxy (e.g., Lehnert \& Heckman 1996), a sample of local galaxies
studied in both 21 cm neutral hydrogen and optical emission lines shows
generally good agreement (Kobulnicky \& Gebhardt 2000; Telles \&
Terlevich 1993).  Taken at face value, these data indicate a significant
evolution in the luminosity-metallicity relation at early epochs.
However, it remains to be established whether the Lyman-drop galaxies
observed to date are representative of the galaxy population at large.
Due to the sensitivity constraints, only the brightest objects have
been observed.  Perhaps, like local HII galaxies, the luminous Lyman
drop galaxies are more luminous than local galaxies of similar
linewidth simply because they are undergoing particularly strong,
concentrated bursts of star formation with amplitudes exceeding those
in any local galaxies.  With typical blue luminosities of $M_B=-22.5$,
these objects are more luminous than any known star-forming (non-AGN)
objects in the local universe, so a suitable low-redshift comparison
sample is not available.

\subsection{Evolution of Chemical Properties}

One general result of this paper is that the probable oxygen abundance
of normal, but very luminous, star-forming galaxies at redshifts near
$z=3$ is subsolar by 0.2-0.6 dex, where the most recent solar
photospheric value is 12+log(O/H)=8.83 (Grevesse \& Sauval 1998).
However, caution must be exercised when relating the oxygen abundance
derived from these empirical emission methods to the solar oxygen
abundance which is determined by meteoritic and solar studies.  The
zero-point of the empirical emission line calibration is subject to an
uncertainty of $\sim0.2$ dex relative to the solar photospheric
measurement.  However, {\it relative} abundance measurements comparing
the high-redshift sample to local galaxies using similar emission line
techniques are still interesting and are more robust.

With these data, we can begin to
study the evolution of another fundamental
galaxy scaling relation, that between luminosity and metallicity
(e.g.,  Lequeux \etal\ 1979; French 1980; Faber 1973; Brodie \& Huchra
1991; Skillman, Kennicutt, \& Hodge 1989; Zaritsky, Kennicutt, \&
Huchra 1994; Richer \& McCall 1995).  The high-redshift sample here is
best compared to local galaxies where metallicities are measured using
the same empirical nebular line method.  Figure~11 shows the
correlation between local irregular and spiral galaxy oxygen abundances
(crosses) from the compilation of Kobulnicky \& Zaritsky (1999) and the
high-redshift sample from Table~4 (filled circles).  Solid squares
denote local H~II galaxies from the sample of Telles \& Terlevich
(1997; magnitudes corrected to the adopted cosmology).  We also plot in
Figure~11 the $z\sim0.4$ sample of star forming galaxies from
Kobulnicky \& Zaritsky (filled triangles).  The $z\sim0.4$ sample is
statistically indistinguishable from the local galaxies.  It is
immediately obvious that the high-redshift sample is 2-4 magnitudes
more luminous for their metallicity than either the local objects or the
$z\sim0.4$ sample.  Objects at $z\sim3$ deviate strongly from the local
luminosity-metallicity relation in a manner similar to local H~II
galaxies.  As noted by Kobulnicky \& Zaritsky (1999), local H~II
galaxies from the sample of Telles \& Terlevich (1997) appear $\sim$1-2
magnitudes more luminous for their metallicity for reasons not yet
identified.   It seems reasonable to propose that both the
high-redshift sample and local H~II galaxies are related in that they
are both undergoing strong bursts of star formation which raise their
luminosities above those of local galaxies with similar chemical
composition and linewidth.  It would be interesting to identify a local
sample of star-forming galaxies for comparison to $z=3$, but we are not
aware of any local (non-AGN) sample which is suitably luminous.

Figure~11 illustrates {qualitatively} how the evolution of these high-redshift objects
might proceed {\it if} they are the ancestors to local ellipticals and spiral
bulges (Steidel \etal 1996a,b; Lowenthal \etal\ 1997).  Not only must
they fade by 2-4 magnitudes, but they must also become more chemically
rich by 0.2-0.5 dex in their gas phase oxygen abundance.  (We note in the 
next section that the ISM metallicity in the high-z sample is
consistent with the metallicity of bulge stars and metal-rich globular
clusters in the Milky Way.)  The solid arrow shows the direction of
evolution if a combination of fading and continued metal enrichment are
both dominant influences.  The dashed arrow shows qualitatively the
evolution if metal enrichment is suppressed by continued inflow of
metal-poor gas as fading proceeds.

Figure~12 shows the correlation between oxygen abundance and kinematic
velocity width, $\sigma_v$, for the same samples of objects plotted in
Figure~11.     Kinematic widths for local spiral and irregular
galaxies, corrected for inclination, come from single-dish 21-cm
measurements.  Kinematic widths for the high and intermediate redshift
sample, and for H~II galaxies from Telles \etal\ (1997) are measured
from Balmer lines with no inclination correction applied.  All classes
of objects are consistent with a well-defined relation between
metallicity and kinematic width, except the $z\sim0.4$ sample of
star-forming galaxies from Kobulnicky \& Zaritsky (1999).  This
intermediate redshift sample exhibits a smaller kinematic width than
local galaxies of similar metallicity.  However, since inclination
corrections have not been applied, one plausible explanation is that
these $z\sim0.4$ objects are preferentially seen at low inclination.  A
larger sample of $z\sim0.4$ objects with measured linewidths and oxygen
abundances is needed to understand possible selection effects.  Note
that in this diagram, the dispersion in the metallicity-linewidth
relation is much smaller than the dispersion in the
metallicity-luminosity plot of Figure~11.  Presumably the tighter
correlation is because both the oxygen abundance and linewidth are
distance-independent and, to first order, extinction-independent
quantities.  Given the sophistication
of today's galaxy evolution models, it would be worthwhile to compare
the kinematic and chemical data in this figure with the popular
semi-analytic simulations.  At the present time,
we are not aware of any suitable model velocity-metallicity 
tabulations in the literature.  The models presented by
Cole \etal\ (2000) show a dispersion of $\sim1$ dex in metallicity for
a given B-band luminosity (their Figure~10), suggesting that the 
dispersion in the model metallicities is larger than the dispersion
observed in real galaxies.

\subsection{Extended [O~III] Emission in Lynx~2-9691}

The spatially extended nature of the [O~III] $\lambda$5007 line in
Lynx~2-9691 is noteworthy in that we detect emission over a distance of
$\sim30$ kpc.  The line exhibits no detectable velocity shift ($<$120
\kms) along this extent.  Not only does this extended morphology rule
out a central AGN as the ionizing source, but it also means that hard
photons capable of ionizing $O^+$ are distributed over a large region
comparable to the size of the present day Milky Way.  Although not as
large as the $100h^{-1}$ kpc Lyman $\alpha$ nebulae discovered around Lyman
drop galaxies (Steidel \etal\ 2000), the physical origin may be the
same.  Steidel \etal\ proposed that the Lyman $\alpha$ nebulae are produced
either by star formation (which is heavily obscured along our line of
sight so that the stellar continuum is not observed) or within cooling
flows.  However, the large velocity spread of 2000 \kms\ in the Steidel
\etal\ objects is not observed here.  Since the stellar continuum is
not detected in the IR spectra of
Lynx~2-9691, we cannot address the spatial distribution
of the nebulae with respect to the galaxy, except to note that the
direct image of the object appears starlike.  Examination of 2-D LRIS
spectra in which we detect the Lyman $\alpha$ line (Figure~1b)
shows a stellar continuum size
consistent with the seeing disk of 0.9\arcsec\ FWHM. The Lyman $\alpha$
emission is only slightly more extended with a FWHM of 1.1\arcsec.
The [O~III] emission seen in the IR spectra is clearly much more
extended than either the UV stellar continuum or the Lyman $\alpha$
emission.  However, a direct comparison at the same position
angle is not possible with the current data.  

As mentioned previously, it is possible that Lynx~2-9691 is lensed by
the $z=0.31$ early type galaxy Lynx~2-9689.  If so, the
spatially-extended [O~III] seen in the NIRSPEC slit might be part of the
lensed arc while the LRIS slit which crosses perpendicular to the arc
would yield only a small emission region.  Simple lensing
models show that the $z=0.31$ galaxy needs to
have a velocity dispersion of $\geq300$ \kms\ to serve as a lens that
amplifies the $z=2.9$ galaxy by several times along the arc and still be
compatible with the 3.5\arcsec\ separation.  

\subsection{The Descendants of Lyman Break Galaxies}

Gas phase oxygen abundances of Lyman-drop galaxies can provide
metallicity estimates for the underlying stellar population.  The warm
ionized component of the ISM in local starbursts is chemically
homogeneous on scales of several kpc and shows no signs of localized
chemical enrichment in the immediate vicinity of the starburst
(Kobulnicky \& Skillman 1996; 1997).  Metals from massive starbursts
may be entrained within the hot interior of wind-driven superbubbles
and be vented into galactic halos before cooling and mixing with the
neutral and molecular gas (Tenorio-Tagle 1996; MacLow \& Ferrara
1999).  Thus, it is reasonable to assume that {\it the recently-formed
massive stars in Lyman-drop galaxies have metallicities very similar to
that of the warm ISM, between 0.25 $Z_\odot$ and 0.9 $Z_\odot$.}
Corroborating evidence comes from Pettini \etal\ (2000) who conclude
that the stellar and interstellar absorption features in MS 1512-cB58
indicate a metallicity of $Z\sim0.25 Z_\odot$ for the massive stars and
intervening cool gas.   Teplitz \etal\ find a similar metallicity
($Z\sim0.3 Z_\odot$) from the oxygen abundance of the warm ionized gas
(Teplitz \etal\ 2000b).

If the intense star formation activity seen in Lyman-drop galaxies
represents the dominant star formation episodes for today's spiral
bulges or giant ellipticals, then the evolved descendants in the local
universe should exhibit similarly subsolar metallicities in their
dominant stellar populations which formed 8-10 Gyr ago.  However, if
the Lyman-drop galaxies are the building blocks which undergo a series
of mergers to form present-day bulges and ellipticals (Lowenthal
\etal\ 1997), then any relationship between stellar age and metallicity
would be substantially more complicated.  In hierarchical galaxy
formation scenarios (e.g., White \& Rees 1978; Searle \& Zinn 1978;
Blumenthal \etal\ 1984; Kauffman \etal\ 1993; Cole \etal\ 1984;
Somerville \etal\ 2000), galaxies grow by mergers of dark matter halos
and accumulation of sub-galactic fragments.  These types of models
predict large dispersions in the age-metallicity relations.  Such large
variations are actually observed in Galactic halo stars and Galactic
globular clusters (Carney 1996; C\^ot\'e \etal\ 2000).  The large
dispersion observed in the age-metallicity relationship of Milky Way
stars suggests a rather complex evolutionary scenario (Edvardsson
\etal\ 1993; Friel \& Janes 1993) for the Galactic disk as well.

It is interesting to compare the ISM metallicities of the $z\sim3$
Lyman-drop objects in Table~4 ($Z = 0.25-0.95~Z_\odot$) with the mean
metallicities of damped Lyman $\alpha$ systems at similar redshifts.
Pettini \etal\ (1997) and Lu \etal\ (1996) derive much lower
metallicities, between 1/300 $Z_\odot$ and 1/10 $Z_\odot$, in
absorption line systems at similar redshifts.  The straightforward
interpretation of this metallicity difference is that damped Lyman
$\alpha$ measurements sample a different population of objects, or at
least a very different physical volume, than the luminous Lyman-drop
galaxies which harbor most of the star formation activity 
(see Pettini \etal\ 1998).

It is also interesting to compare the mean ISM metallicity of the
objects in Table~4 ($Z\sim0.3 Z_\odot ~ \equiv~ [O/H]\sim-0.4\pm0.3$) to the mean
metallicity of Galactic bulge stars and metal-rich Galactic globular
clusters summarized in C\^ot\'e \etal\ (2000): $[Fe/H]\sim-0.3$ for
bulge stars and  $[Fe/H]\sim-0.6$ for metal-rich Galactic globular
cluster.  Outside the Milky Way, in giant ellipticals, the metallicity
of the metal-rich globular clusters is known to correlate with the mass
of the parent galaxy (Forbes, Brodie, \& Grillmair 1997), varying
between $-0.4<[Fe/H]<0.0$.  Evidently the formation and evolution of
these metal-rich globular clusters is coupled to that of the parent
galaxy and they represent the galaxy's intrinsic cluster population
(C\^ot\'e \etal\ 1998).  Assuming $[Fe/O]\sim$0.0, as is typical for
Galactic stars with $[Fe/H]>-1$ (Wheeler, Sneden \& Truran 1989), we
can compare the metallicities of the ISM in Lyman-drop galaxies to the
stellar components in nearby galaxies.  {\it The metallicity of the ISM
in the Lyman-drop galaxies tabulated here, $[O/H]\sim-0.4\pm0.3$, is
consistent with the metallicities needed to form the metal-rich
globular cluster population in the Milky Way, $[Fe/H]\sim-0.6\pm0.2$. }
Many of the Galaxy's bulge and disk stars have metallicities and ages
that would be consistent with their formation in a galaxy like
Lynx2-9691 at $z=2.9$ as well.

\section{Conclusions}

The mean metallicities of $z\sim3$ galaxies lie in the range $Z =
0.25-0.95~Z_\odot$, considerably more metal-rich than the mean
damped Lyman $\alpha$ systems observed at the same redshift which have
$Z\leq0.1~Z_\odot$ (Pettini \etal\ 1997; Lu \etal\ 1996).  
Based on the limited data available to date, the population of
star-forming galaxies at $\sim3$ is found to be 2-4 magnitudes more
luminous than galaxies of similar rotational velocity and metallicity
today.  If these galaxies are representative of the populations at
early times, then the fundamental galaxy scaling relations, namely the
luminosity-velocity and luminosity-metallicity relation may be
significantly different at $z>2.3$ compared to the present.  If,
however, we are only viewing the peak of the luminosity function,
perhaps the result of being viewed during the earliest phases of a
strong burst of star formation, then these data may merely indicate a
larger dispersion of luminosities at high-redshift compared to today.
The gas-phase metallicities of Lyman-drop galaxies studied here are
comparable to the metallicities needed to form stars in Galactic
metal-rich globular clusters and bulge.  With the arrival of many 8-10
m class telescopes with near-IR spectrographs, we expect that these
early, and still uncertain, results will soon be supplemented by large
enough samples to begin statistical studies of the velocity and
chemical properties at $z>2.3$ at the same significance level as that
of photometric and morphological studies at high-redshift.

\section{Acknowledgments}

We thank the NIRSPEC instrument team for bringing such a successful
instrument to fruition, and the Keck staff for their support,
especially Greg Wirth who assisted with observations at the telescope.
Nicolas Cardiel cleverly solved a problematic flux calibration issue.
Thanks to James Lowenthal for providing optical spectra of the Coup
Fourr\'e galaxy in electronic form and he and Nathan Roche for early
access to their paper on the CFg.  Matt Bershady provided the scanned
plate image of Lynx~2-9691.  We gratefully acknowledge educational
exchanges with Matt Bershady, James Bullock, Dirk Fabian, Jay
Gallagher, Max Pettini, Philipp Richter, Rachel Somerville, Chuck Steidel, and Harry
Teplitz.  This work was supported by a faculty research grant from the
University of California, Santa Cruz, and an NSF grant AST-9529098.
H.~A.~K is grateful for support from Hubble Fellowship grant
\#HF-01094.01-97A awarded by the Space Telescope Science Institute
which is operated by the Association of Universities for Research in
Astronomy, Inc. for NASA under contract NAS 5-26555.

\clearpage
\centerline{\psfig{file=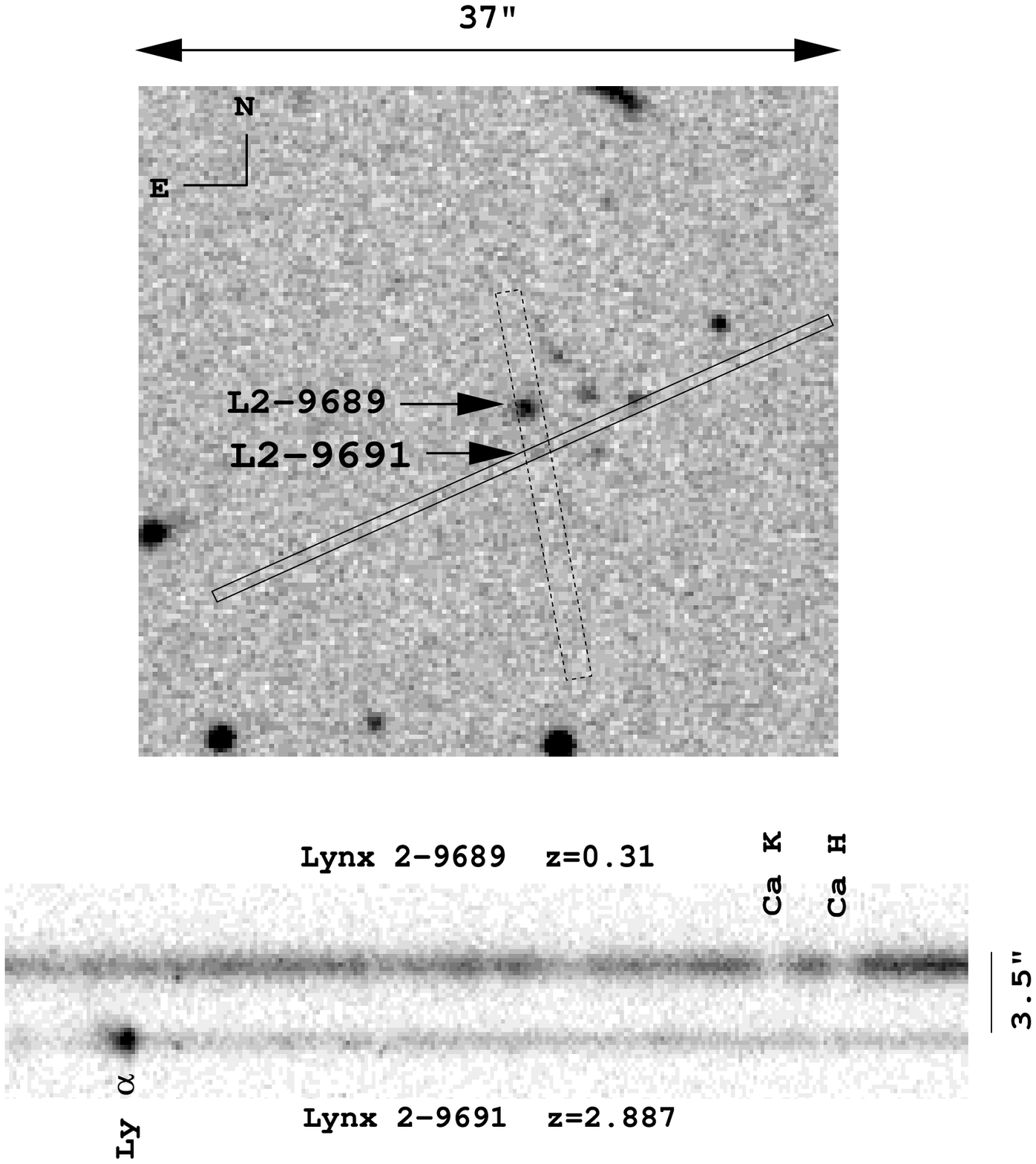,height=6.0in}} \figcaption[f1a.ps]{(a)
Finding chart for Lynx~2-9691 from microdensitometer scans of a
relatively deep 4-m prime focus photographic plate taken with the $R_F$
filter (M. Bershady, private communication).  Table~1 lists 2000 epoch
coordinates.  A solid rectangle marks the position of the NIRSPEC
slit.  A dashed rectangle denotes the position of the optical Keck/LRIS
slitlet.  (b)---Keck/LRIS discovery spectrum of Lynx~2-9691 showing
Lyman $\alpha$ emission at $z=2.8877$.  An early-type galaxy,
Lynx~2-9689, showing prominent Ca H\&K $\lambda\lambda3970,3932$
absorption at $z=0.31$, is located 3.5\arcsec\ to the northeast at a
position angle of 10 degrees.  The spatial scale of the CCD is
0.21\arcsec/pix. \label{fig1a} }

\centerline{\psfig{file=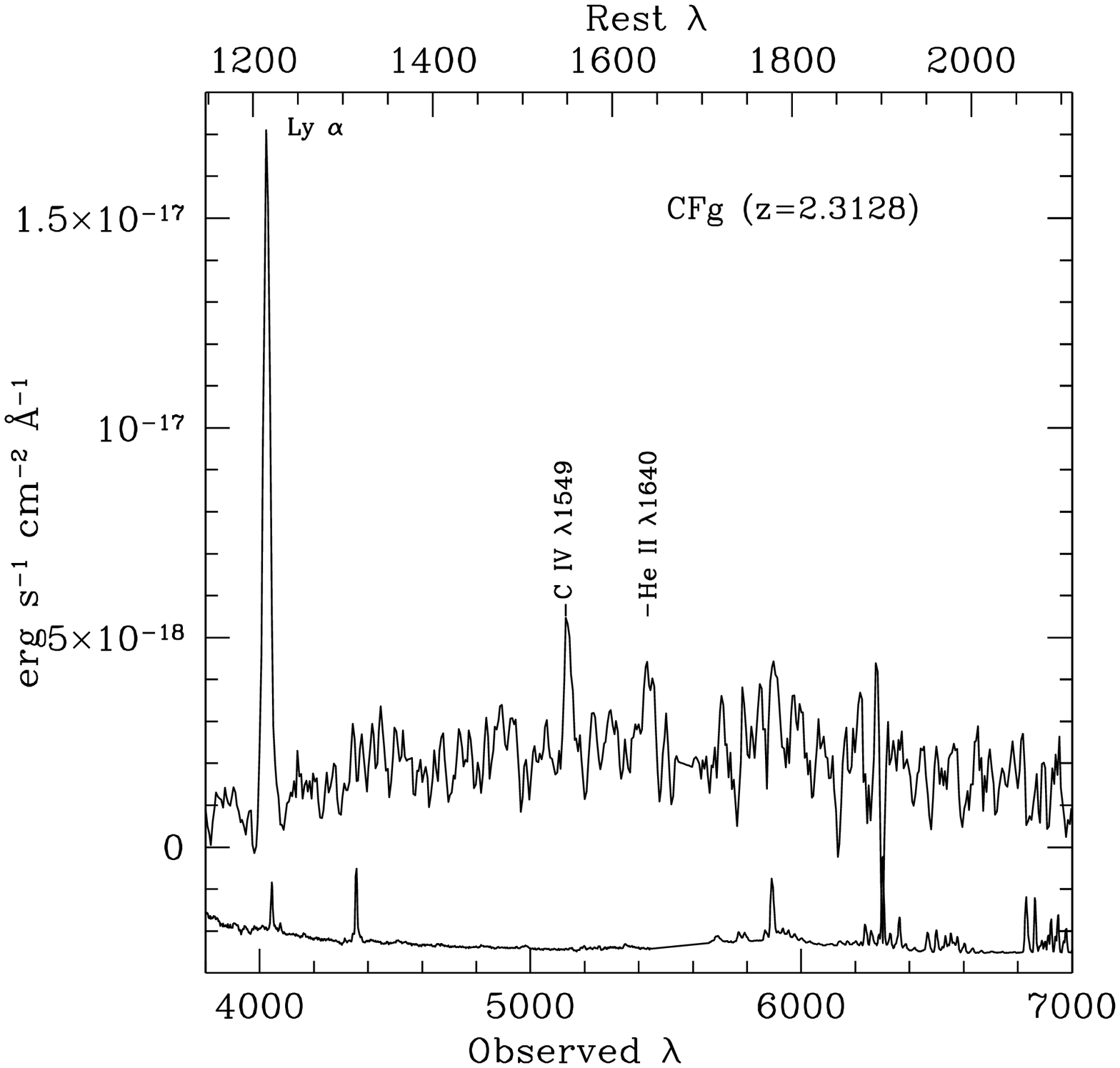,height=6.0in}} \figcaption[f2.ps]{Optical
spectrum of the Coup Fourr\'e Galaxy (CFg) from Lowenthal
\etal\ (1991).  Lyman $\alpha$ and narrow emission lines from C IV
$\lambda$1548/51 and He II $\lambda$1640 are visible.   A
representative night sky airglow  spectrum is shown below, scaled
arbitrarily.  As in Lowenthal \etal\ (1991), the spectrum has been
smoothed to a resolution of FWHM$\simeq$24 \AA.  \label{spec1} }

\centerline{\psfig{file=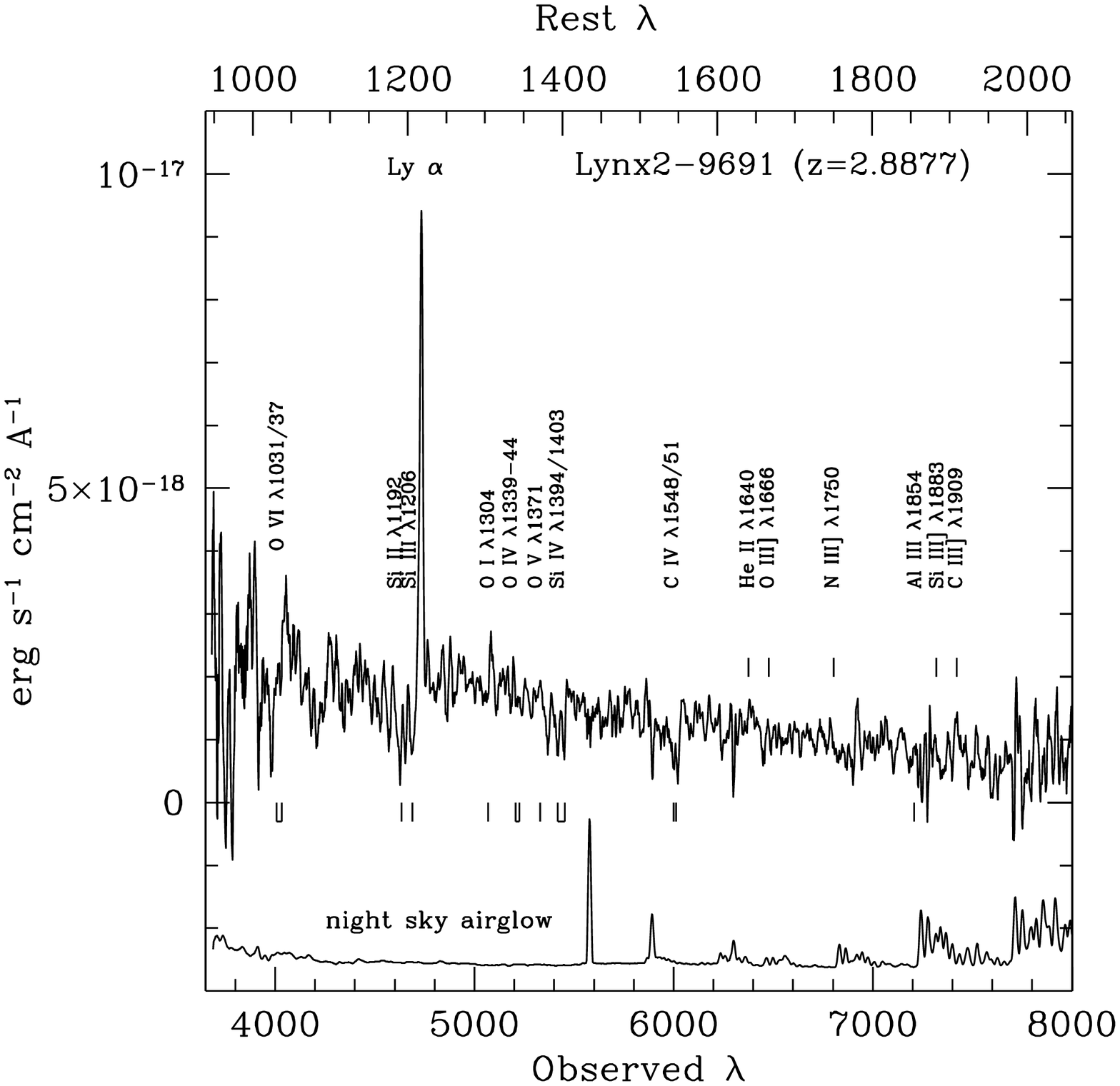,height=5.0in}} \figcaption[f3.ps]{Keck
LRIS optical spectrum of Lynx~2-9691 with a resolution of 9 \AA\ FWHM.
The spectrum has been smoothed with a 5-pixel boxcar for display
purposes. A representative night sky airglow  spectrum is shown below,
scaled arbitrarily.  Common interstellar emission and absorption lines
allow us to derive a redshift of z=2.889 consistent with the z=2.8877
derived from the infrared spectra in Figure 6.  The lack of broad lines
and the preponderance of stellar absorption features, (see also
Figure~4) is consistent with a normal star-forming galaxy rather than
an active nucleus.  As commonly seen in other Lyman-drop galaxies
(Steidel \etal 1996a,b; Pettini \etal\ 2000), the Lyman $\alpha$ centroid
is shifted redward by 5 \AA\ relative to the nominal position expected
for its redshift, suggesting absorption of the blue wing by the
galaxy's intervening ISM.  \label{spec2}}

\centerline{\psfig{file=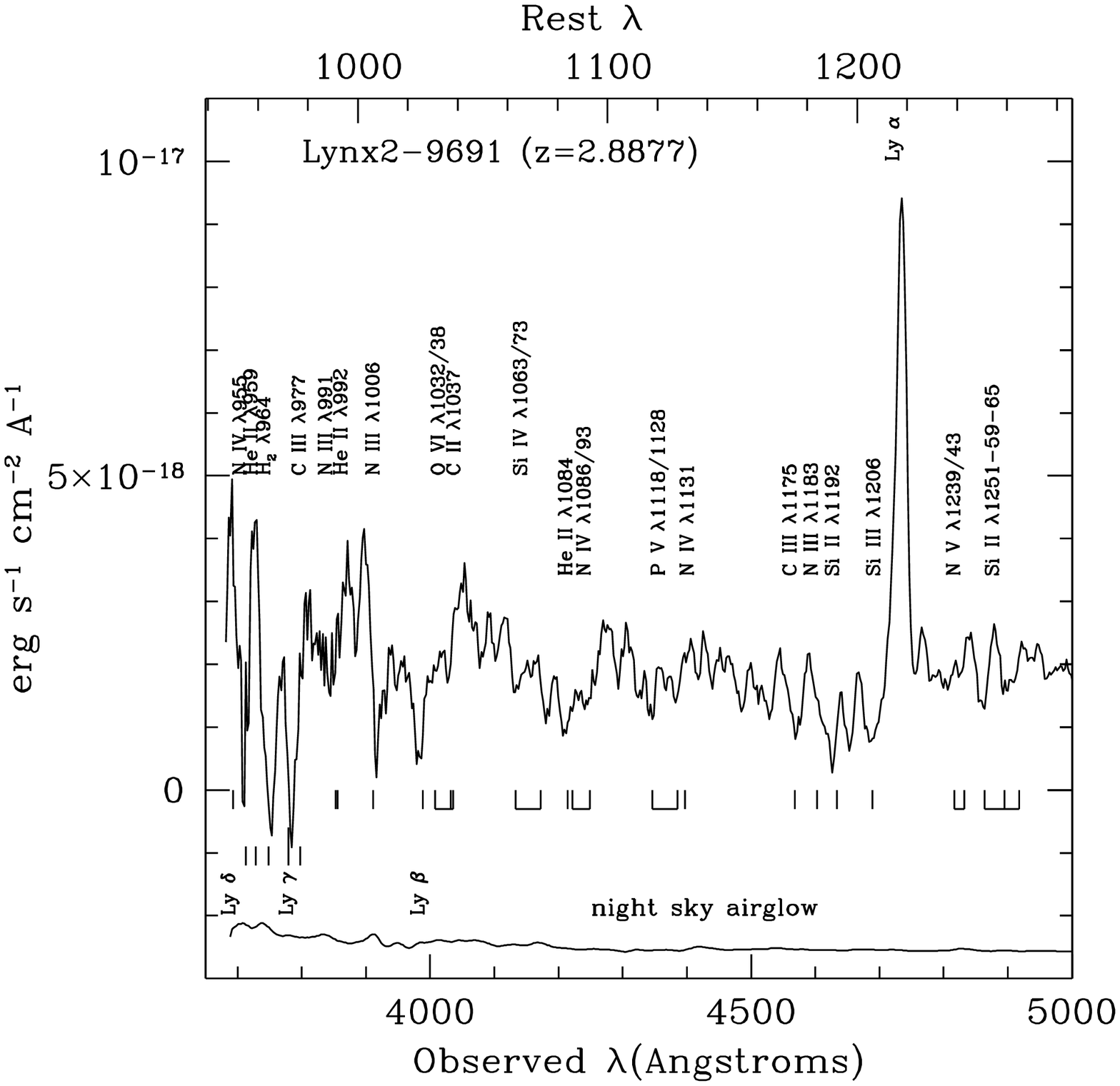,height=5.0in}}
\figcaption[f4.ps]{Enlargement of the far-ultraviolet portion of the
Lynx~2-9691 spectrum.  Strong stellar and interstellar absorption
features dominate the spectrum.  A representative night sky airglow
spectrum is shown below, scaled arbitrarily. \label{spec3}}

\centerline{\psfig{file=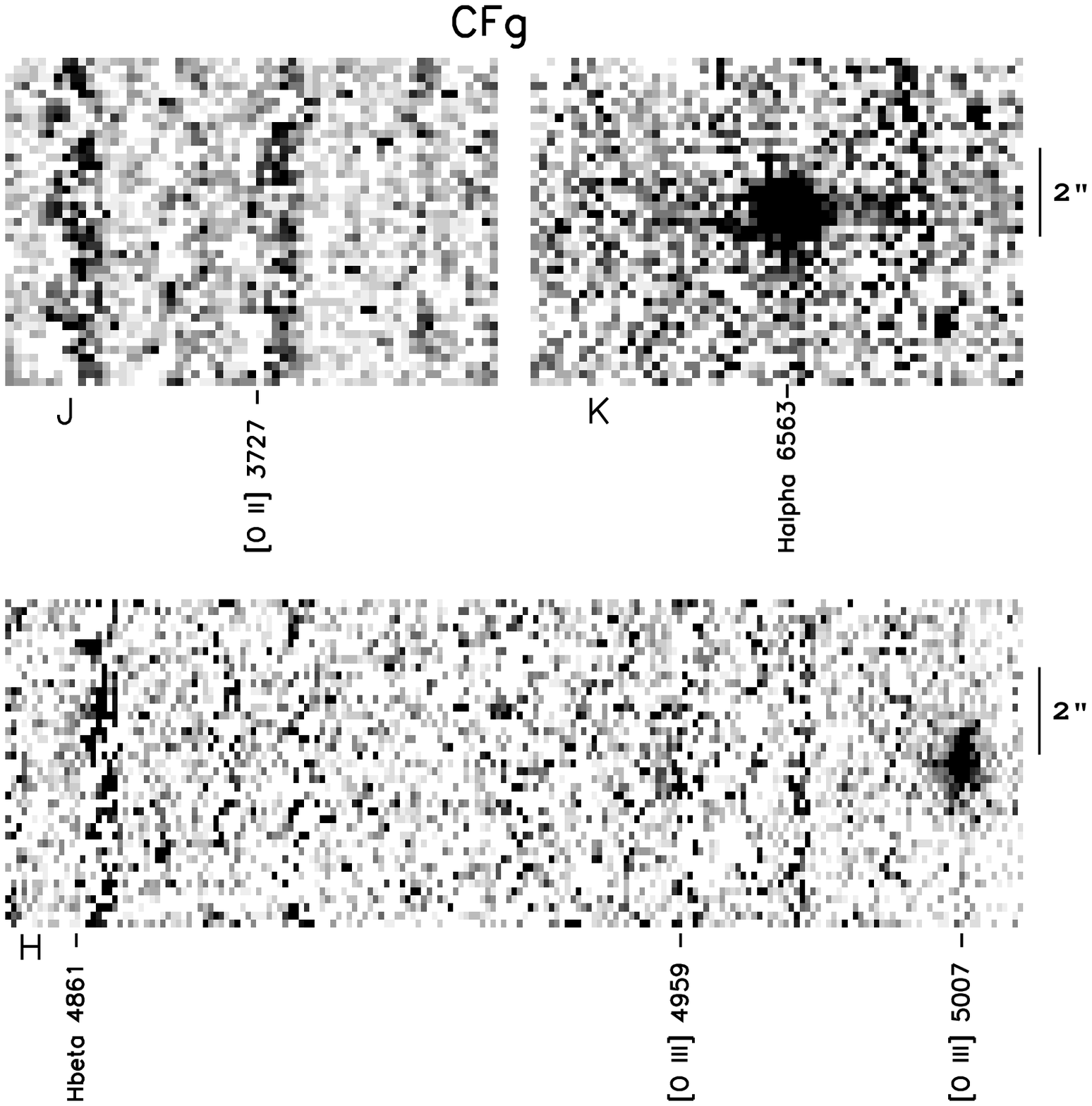,height=6.0in} }
\figcaption[f5.ps]{Two-dimensional sub-images of the
Keck NIRSPEC spectra of the CFg in the J, H, and K infrared bandpasses.
Positions of strong nebular emission lines are labeled.  The scale is
0.178\arcsec/pix in the spatial direction and $\sim3$ \AA/pix in the
spectral direction.  Labels mark the expected positions of strong
nebular lines assuming $z=2.3128$.  \label{spec4}}

\centerline{\psfig{file=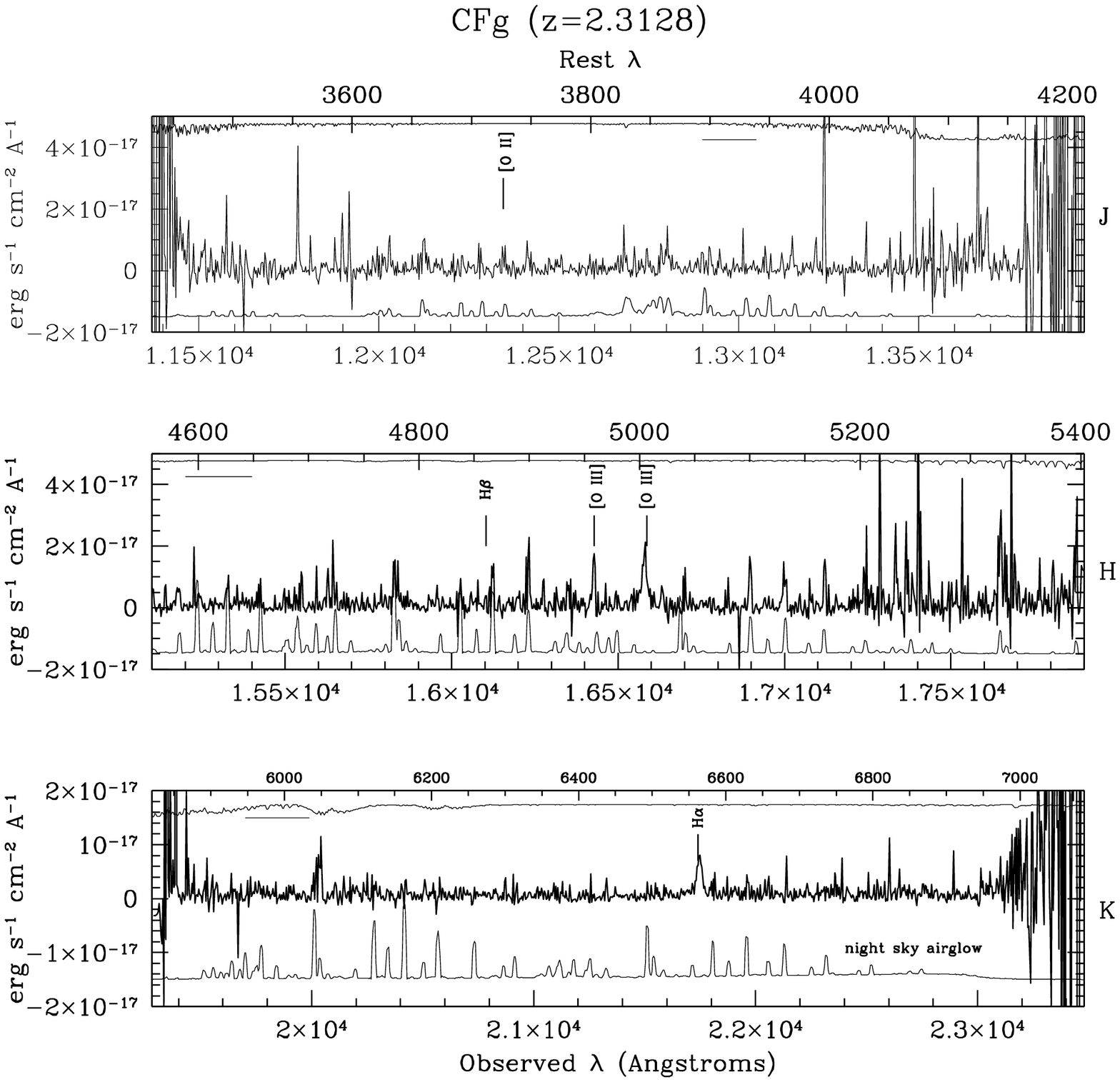,height=6.0in}} \figcaption[f6.ps]{Keck
NIRSPEC infrared spectra of the CFg in the J,H,K bandpasses.  The
positions of strong nebular emission lines are marked assuming
$z=2.3128$.  For reference, a spectrum of the night sky airglow is
plotted below, scaled arbitrarily.  The atmospheric transmissivity
curve based on models for Mauna Kea (Lord 1992) is plotted above each
spectrum, scaled arbitrarily, with a horizontal scale bar indicating
zero transmissivity.  \label{spec5}}

\centerline{\psfig{file=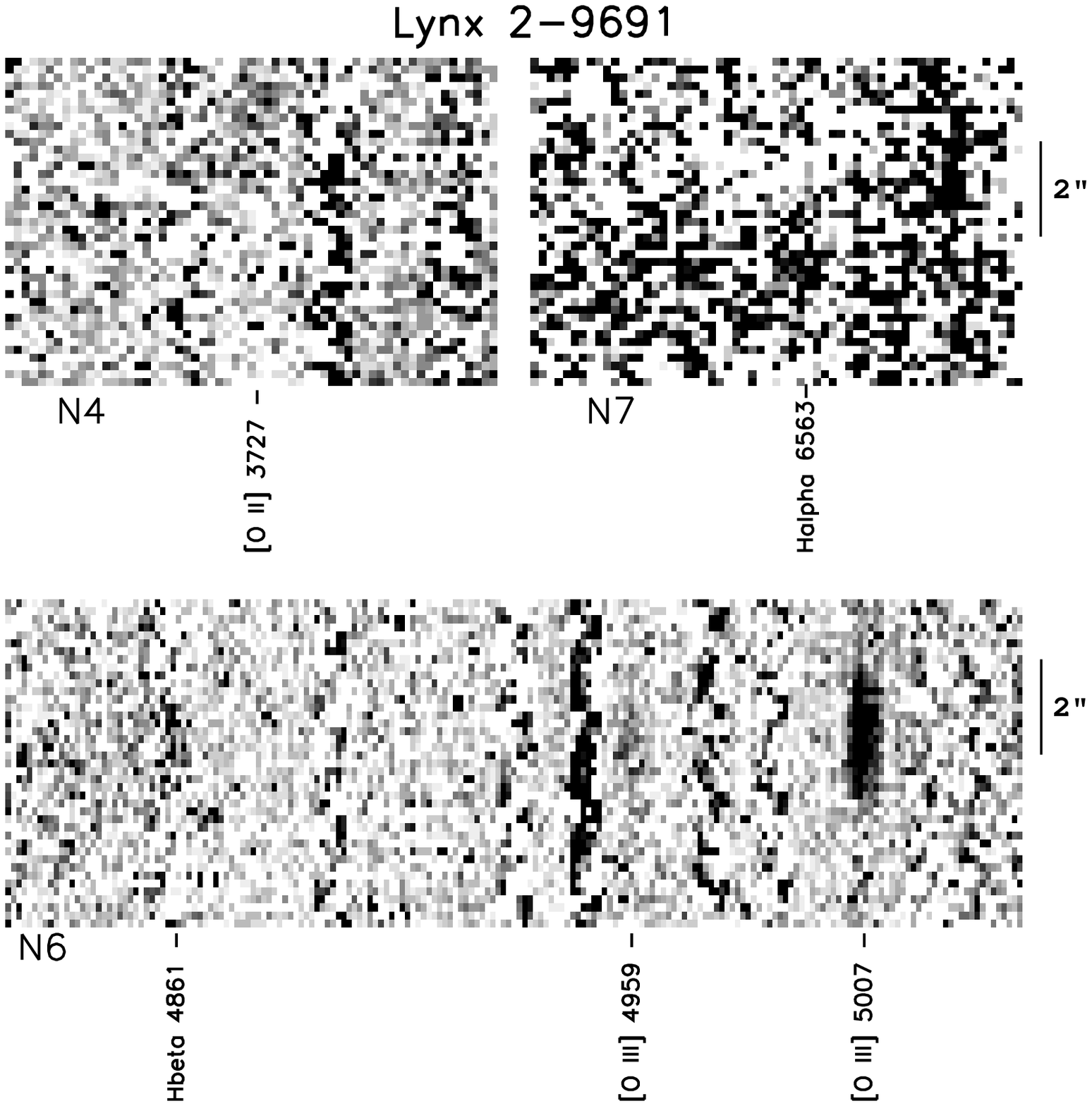,height=6.0in} }
\figcaption[f7.ps]{Two-dimensional sub-images of the Keck NIRSPEC
spectra of Lynx~2-9691 in the NIRSPEC N4, N6, N7 infrared bandpasses.
Positions of strong nebular emission lines are labeled assuming
$z=2.8877$.  The scale is 0.178\arcsec/pix in the spatial direction and
$\sim3.8$ \AA/pix in the spectral direction.  Labels mark the expected
positions of strong nebular lines.  \label{spec6}}

\centerline{\psfig{file=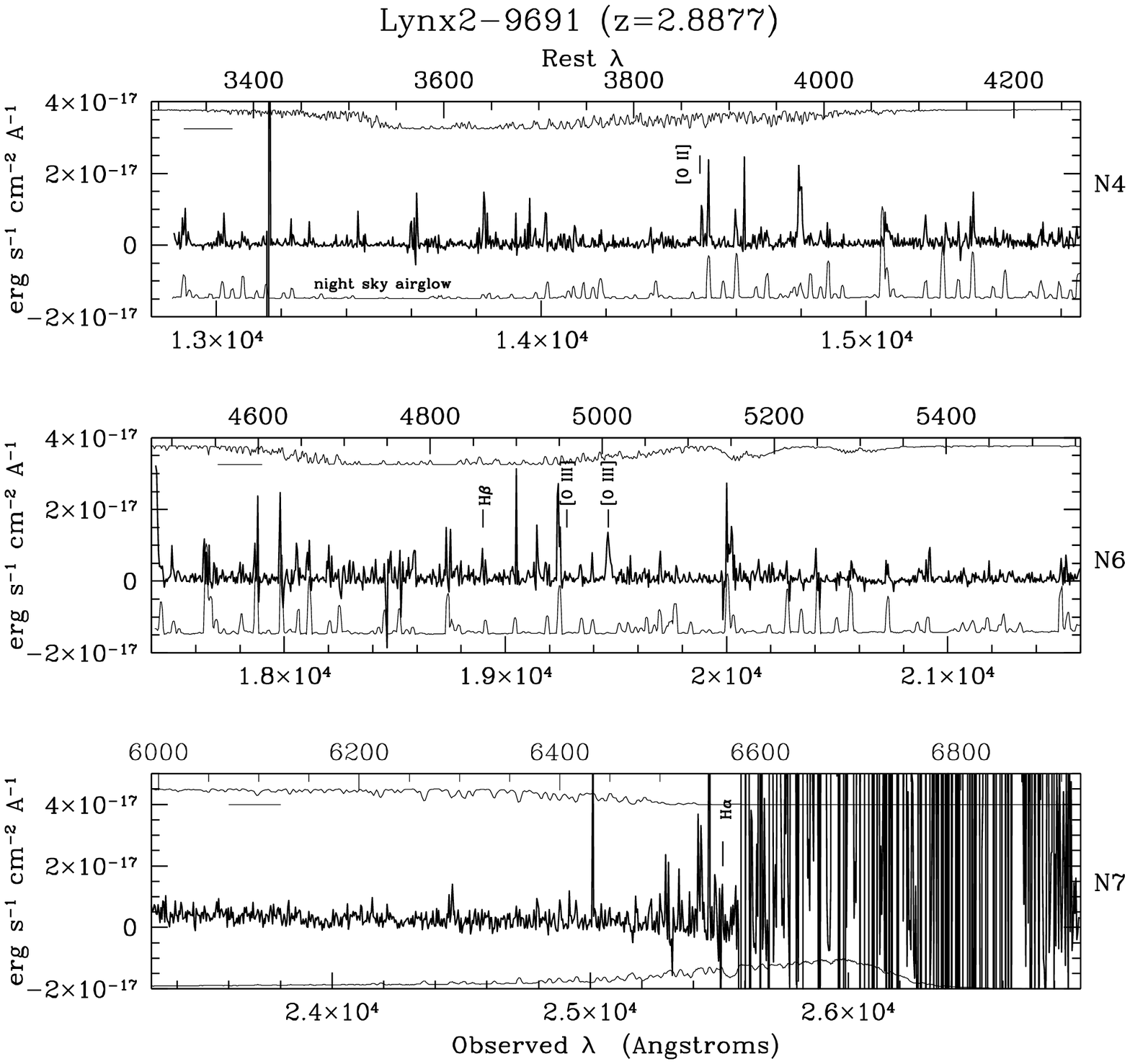,height=6.0in}} \figcaption[f8.ps]{NIRSPEC
infrared spectra of the CFg in the NIRSPEC N4,N6,N7 bandpasses.  The
positions of strong nebular emission lines are marked assuming
$z=2.8877$.  For reference, a spectrum of the night sky airglow is
plotted below, scaled arbitrarily.   The atmospheric transmissivity
curve based on models for Mauna Kea (Lord 1992) is plotted above each
spectrum, scaled arbitrarily, with a horizontal scale bar indicating
zero transmissivity. The spectrum becomes noisy beyond
2.5 $\mu$m because of thermal background emission.  \label{spec7}}

\centerline{\psfig{file=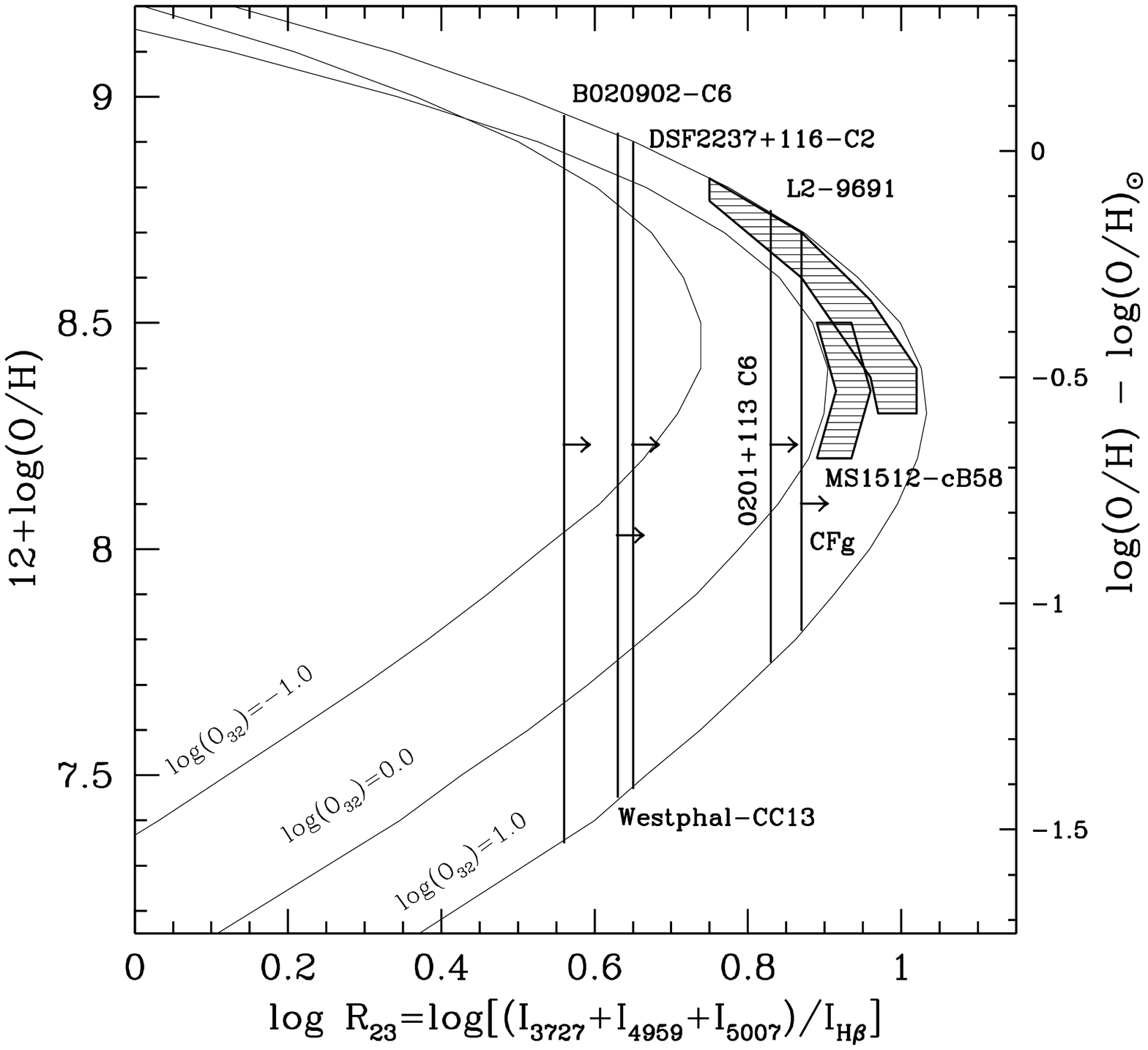,height=6.0in}} \figcaption[f9.ps]{The
diagnostic line ratio $R_{23}=[(I_{3727}+I_{4959}+I_{5007})/I_{H\beta}$
for high-redshift objects from the literature, and the corresponding
oxygen abundance, 12+log(O/H), from the models of McGaugh (1991).
Shaded boxes represent the most probable locations for the
best-measured objects, Lynx~2-9691 and MS1512-cB58 (Teplitz
\etal\ 2000b).  Other objects with $R_{23}$ lower limits are plotted
spanning a range of possible metallicities.  \label{R23diag}}

\centerline{\psfig{file=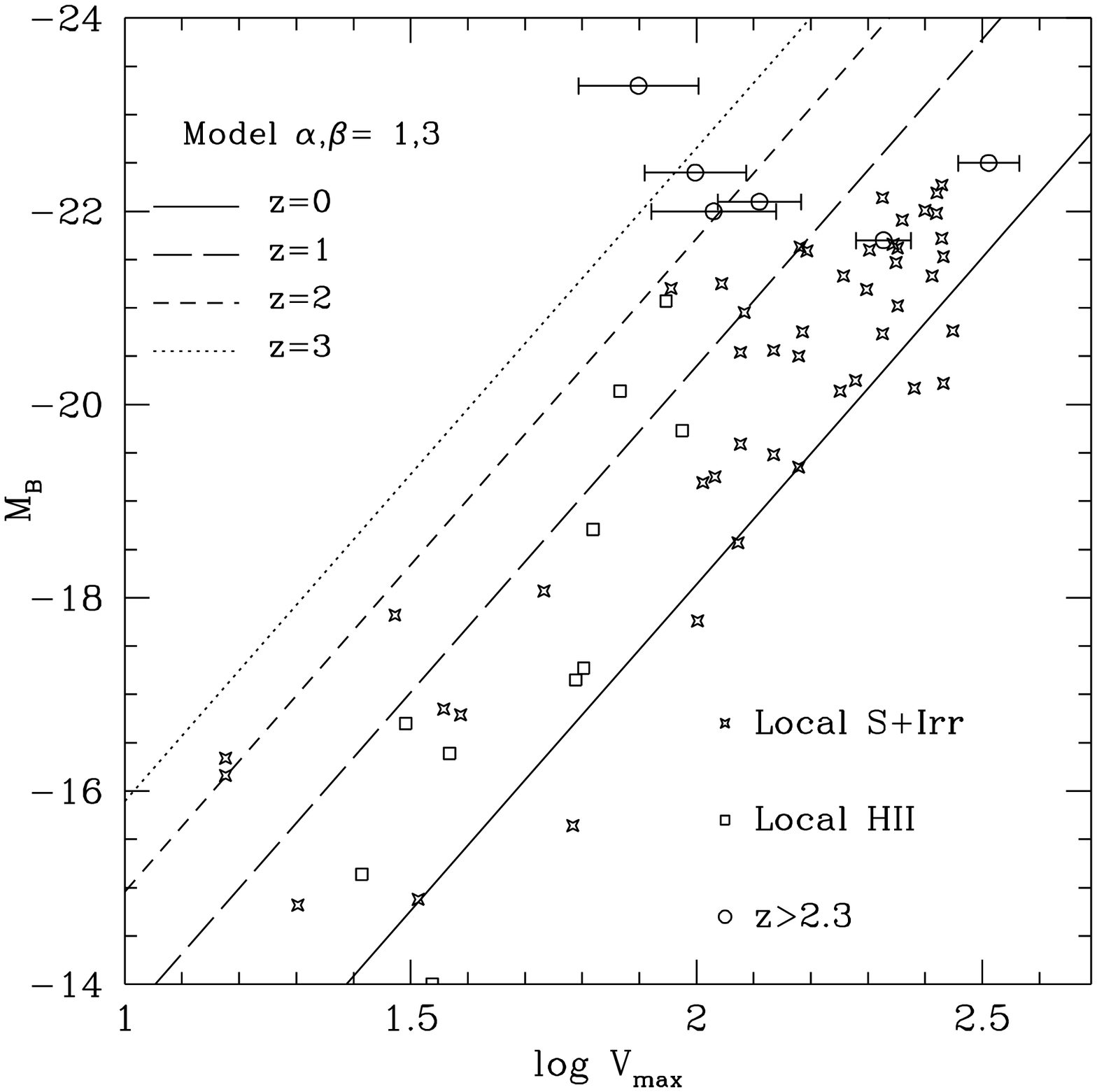,height=6.0in}}
\figcaption[f10.ps]{Evolution of the relation between luminosity,
$M_B$, and circular velocity, $v_m$, as a function of redshift.  The
figure illustrates a range of possible evolution of the luminosity-velocity
relation as a
function of the $\alpha$,$\beta$ parameters of Bullock \etal\ (2000;
see Equation 5).  We show data for local
spiral and irregular galaxies compiled in Kobulnicky \& Zaritsky
(1999).  High-redshift objects from Table~4 appear as open circles
with error bars, while local HII galaxies from Telles \& Terlevich (1997) appear as open
squares.  Like local HII galaxies, $z\sim3$ Lyman-drop galaxies are
systematically more luminous for their velocity widths than
spiral and irregular galaxies in the local luminosity-linewidth
correlation. \label{V_MB}}

\centerline{\psfig{file=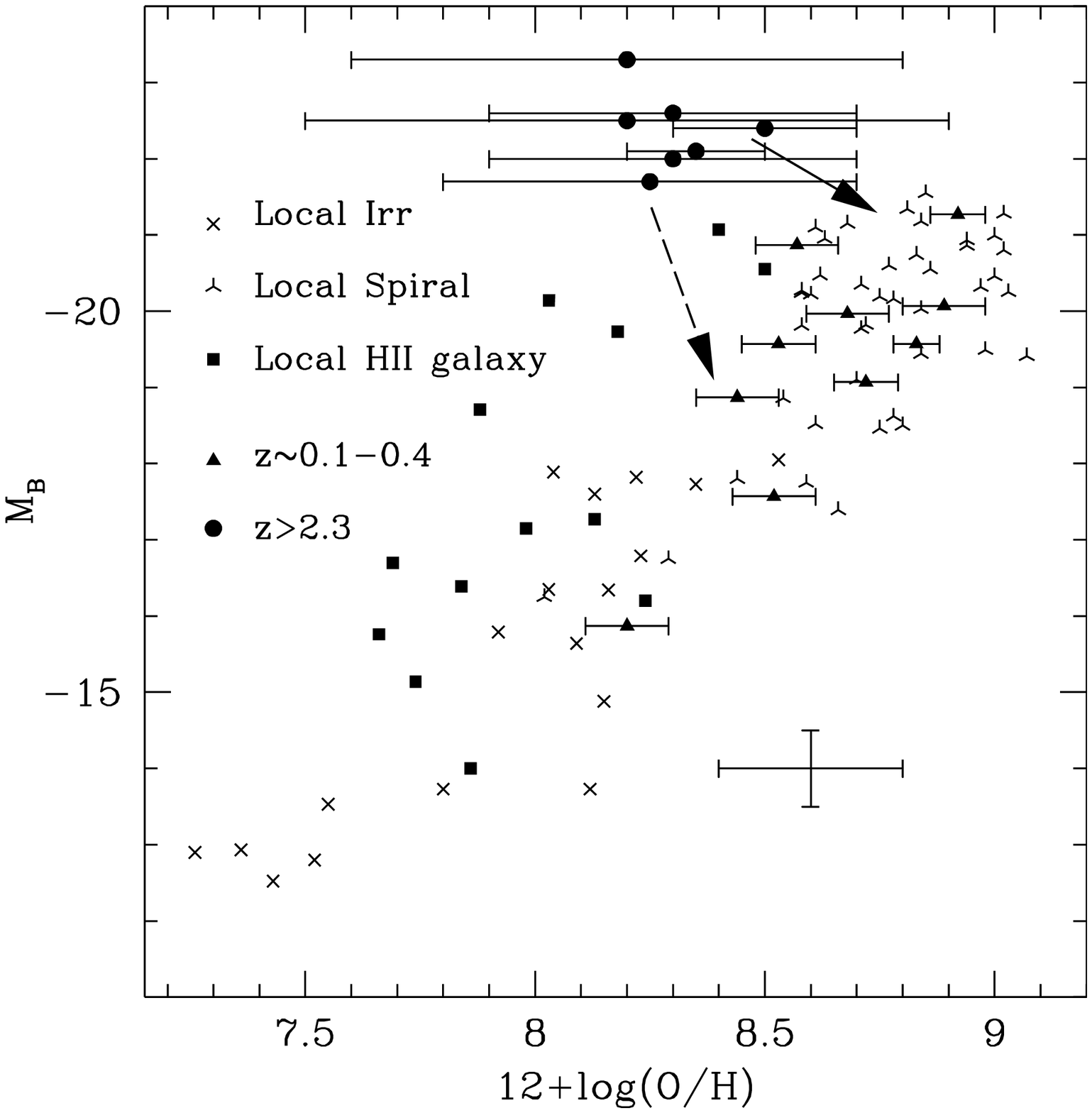,height=5.0in}}
\figcaption[f11.ps]{Oxygen abundance, 12+log(O/H), versus the absolute
restframe B magnitude.  Local irregular and spiral galaxies collected
from the literature and Kobulnicky, Kennicutt, \& Pizagno (1999) appear
as small skeletal symbols.  A representative error bar appears in the
lower right.  The sample of intermediate-redshift $z\sim0.4$  emission
line galaxies from Kobulnicky \& Zaritsky (1999) appears as solid
triangles.  Solid squares denote local H~II galaxies from the sample of
Telles \& Terlevich (1997; magnitudes corrected to the adopted
cosmology).  High-redshift objects from Table~4 appear as filled
circles with error bars.   They are significantly more luminous for
their metallicity than local objects.  Under the assumption that
Lyman-drop galaxies evolve into modern day ellipticals and galactic
spheroids, arrows indicate the tracks of high-redshift galaxies as they
fade and become more chemically enriched.  The solid arrow shows
qualitatively the path of a galaxy which fades as its interstellar
medium becomes more chemically rich by the continued cycles of star
formation and metal production.  The dashed line shows qualitatively
the evolutionary path if continued infall of metal-poor gas is a large
effect as fading proceeds.  \label{fig11}}

\centerline{\psfig{file=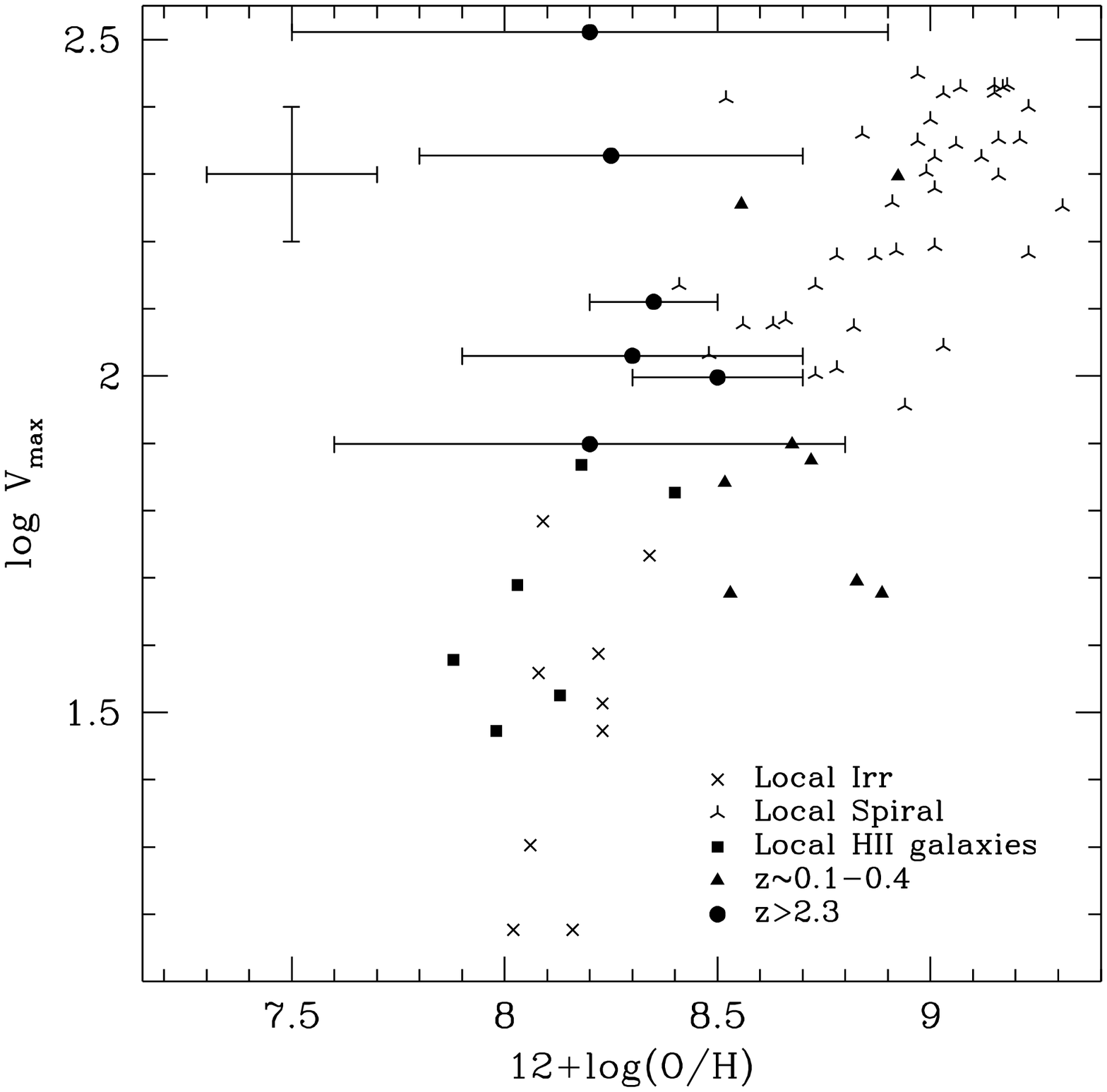,height=6.0in}} \figcaption[f12.ps]{
Oxygen abundance, 12+log(O/H), versus velocity dispersion, $\sigma_v$.
A representative error bar appears in the lower right. Symbols are the
same as in Figure~11.   Kinematic widths for local spiral and irregular
galaxies, corrected for inclination, come from single-dish 21-cm
measurements.  Kinematic widths for the high and intermediate redshift
sample and for H~II galaxies from Telles \etal\ (1997) are measured
from Balmer lines with no inclination correction applied, and thus,
strictly speaking, are lower limits.  The $z\sim3$ data are marginally 
consistent with the local
metallicity-linewidth relation.  Note that HII galaxies and the
$z\sim3$ sample show better agreement with local spirals and irregulars than in Figure~10.
  \label{VOH}}

\begin{deluxetable}{lrcrc}  
\tablecolumns{5}  
\tablewidth{0pc}  
\tablecaption{Target Objects}  
\tablehead{  
\colhead{Name}         &\colhead{CFg } & \colhead{Comment} &\colhead{Lynx~2-9691} & \colhead{Comment}   }
\startdata  
RA (2000)              & 01 03 08.43    & A                 & 08 43 31.88          & B \\
DEC (2000)             & +13 16 39.6    & A                 & +44 44 31.2          & B \\
Redshift (z)           & 2.3128         & A                 & 2.8877               & B \\
$B_J$\tablenotemark{1}                  & \nodata        & \nodata           & 23.88               & B \\
$R_F$\tablenotemark{1}                  & \nodata        & \nodata           & 23.04               & B \\
$I_{814}$              & 23.85          & A                 & \nodata              & \nodata \\
$V_{555}$              & 24.14          & A                 & \nodata              & \nodata \\
$r_{hl} (kpc)$         & 5.7            & A                 & 15                 & C \\
\enddata  
\tablerefs{
(A) Roche et. al 2000;
(B) This paper;
(C) This is the radius of the [O~III] emission reported here.
}
\tablenotetext{1}{Photographic magnitudes on the optical $JK$ system of Kron (1980).
See also Munn \etal\ (1997).}
\end{deluxetable}

\begin{deluxetable}{lrr}  
\tablecolumns{3}  
\tablewidth{0pc}  
\tablecaption{Keck II NIRSPEC Observations}  
\tablehead{  
\colhead{Name}&\colhead{CFgA } &\colhead{Lynx~2-9691}      }
\startdata  
Filter for [O~II]            & J           & N4  \\
\ \ observed $\lambda$ (\AA) & 11370-13960 & 12870-15650 \\
\ \ rest $\lambda$ (\AA)     & 3432-4214   & 3310-4025   \\
\ \ Exposure (s)             & 3x600       & 2x900 \\
\ \ Airmass                  & 1.55        & 1.12  \\
Filter for H$\beta$, [O~III] & H           & N6  \\
\ \ observed $\lambda$ (\AA) & 15100-17900 & 17460-21600  \\
\ \ rest $\lambda$ (\AA)     & 4558-5403   & 4491-5556    \\
\ \ Exposure (s)             & 5x600       & 3x900 \\
\ \ Airmass                  & 1.35        &1.25 \\
Filter for H$\alpha$         & K           & N7  \\     
\ \ observed $\lambda$ (\AA) & 19280-23480 & 23300-26900  \\
\ \ rest $\lambda$ (\AA)     & 5820-7087   & 5993-6919   \\
\ \ Exposure (s)             & 2x900       & 5x450 \\
\ \ Airmass                  & 1.20        & 1.30 \\
\enddata  
\end{deluxetable}  

\begin{deluxetable}{lrcrc}  
\tablecolumns{5}  
\tablewidth{0pc}  
\tablecaption{Spectral Measurements}  
\tablehead{  
\colhead{Emission Line Fluxes}&\colhead{CFg} &\colhead{Comment} &\colhead{Lynx~2-9691} &\colhead{Comment}      }
\startdata  
Ly$\alpha$   & & & & \\
\ \ F ($\times10^{-16}~erg~s^{-1}~cm^{-2}$)           &  5.6          & A         & 1.35$\pm$0.08 & B  \\
\ \ EW (rest frame)                                   &  140          & A         & 18$\pm$2      & B  \\
\ \ FWHM (\AA) (observed)                             &   9.4         & A         & 14.8          & B  \\
$\left[O~II\right]$ $\lambda$3727   & & & & \\
\ \ F ($\times10^{-16}~erg~s^{-1}~cm^{-2}$)           &  $<$0.47        & C       & $<$0.50       & C  \\
\ \ FWHM (\AA) (observed)                             &  \nodata        & \nodata & \nodata       & \nodata  \\
H$\beta$ $\lambda$4861  & & & &  \\
\ \ F ($\times10^{-16}~erg~s^{-1}~cm^{-2}$)           &  0.61           & E       & 0.50$\pm0.17$ & B  \\
\ \ FWHM (\AA) (observed)                             &  \nodata        & \nodata & 10.1          & B  \\
$\left[O~III\right]$ $\lambda$4959   & & & & \\
\ \ F ($\times10^{-16}~erg~s^{-1}~cm^{-2}$)           &  1.41$\pm0.10$  & B        & 0.76          & D  \\
\ \ FWHM (\AA) (observed)                             &  9.5            & B        & \nodata       & \nodata \\
$\left[O~III\right]$ $\lambda$5007   & & & & \\
\ \ F ($\times10^{-16}~erg~s^{-1}~cm^{-2})$           &  2.81$\pm0.13$ & B         & 2.21$\pm0.09$ & B  \\
\ \ FWHM (\AA) (observed)                             &  16.3           & B        & 16.9          & B  \\
H$\alpha$ $\lambda$6563   & & & & \\
\ \ F ($\times10^{-16}~erg~s^{-1}~cm^{-2})$           &  1.65$\pm0.07$  & B       & 1.43          & G  \\
\ \ FWHM (\AA) (observed)                             &  25.6           & B       & \nodata       & \nodata  \\
$L_{H\alpha}~(\times10^{40}~erg~s^{-1})$              & 665             & F       & 1250          &  G       \\
$SFR_{Balmer}$ ($M_\odot~yr^{-1}$)                               & 59              & H       & 111           &  H       \\
$W_{FWHM} ~ (km~s^{-1})$                               & 301            & I       & 156           &  J      \\
$\sigma={FWHM}/2.35 ~ (km~s^{-1})$                     & 128            & I       & 66            &  J      \\
$M_{vir}~(10^{10}~M_\odot)$                            & 7.0            & K       & 5.2           &  K      \\
$F_{1500}~(\times10^{-17}~erg~s^{-1}~cm^{-2}~$\AA$^{-1})$         & 0.25           & M            &  0.15         &  M     \\
$L_{1500}~(\times10^{41}~erg~s^{-1}~$\AA$^{-1})$       & 3.3            & L       &  4.3          &  L    \\
$SFR_{1500}~(M_\odot~yr^{-1})$                         & 31            & M       &  41          &  M    \\
\enddata 
\tablerefs{
(A) From Lowenthal et. al (1991);
(B) Measurement by Gaussian fit;
(C) 3$\sigma$ upper limit;
(D) Assuming $F_{5007}/2.9$;
(E) Assuming no extinction and $F_{H\beta}=F_{H\alpha}/2.86$;
(F) From the observed $H\alpha$ flux and the adopted cosmology;
(G) From the observed $H\beta$ flux assuming no extinction,
$F_{H\alpha}=F_{H\beta}\times2.86=1.4\times10^{-16}~erg~s^{-1}~cm^{-2}$; 
(H) $SFR (M_\odot~yr^{-1}) =
8.9\times10^{-42}~{L_{H\alpha}}~(erg~s^{-1})$, Kennicutt (1983);
(I) Intrinsic FWHM of $H\alpha$ after correction for the
instrumental profile, modeled as a Gaussian with FWHM=13.4 \AA;
(J) Intrinsic FWHM of [O~III] $\lambda5007$ after correction for the
instrumental profile, modeled as a Gaussian with FWHM=13.6 \AA;
(K) Virial mass estimates; see text for cautionary notes.
As a fiducial size for Lynx~2-9691, we use one half of
the spatial extent of the [O~III] emission, 14 kpc.
(L) The specific UV continuum flux and luminosity from Lowenthal \etal\
(1991) and our Keck/LRIS spectra.
(M) Star formation rate as determined from the 1500 \AA\ specific luminosity
adopting the prescription of Madau, Pozzetti, \& Dickinson (1998) as described in the text.
}
\end{deluxetable}

\begin{deluxetable}{lccrrrrrrr}  
\tablecolumns{9} 
\tablewidth{0pc}  
\tablecaption{Properties of High-Redshift Star Forming Objects}  
\tablehead{  
\colhead{Object}&\colhead{${\log}(R_{23})$}&\colhead{$log(O_{32})$}&\colhead{log(O/H)}&\colhead{Ref.}  & \colhead{$K_{AB}$}&\colhead{$M_B$}& \colhead{Ref.}  & \colhead{z} & \colhead{SFR}  \\
\colhead{  }    &\colhead{(1)}             &\colhead{(2)}          &\colhead{+12 (3)} &\colhead{(4)}   &\colhead{(5)}  & \colhead{(6)}   & \colhead{(7)} & \colhead{(8)} & \colhead{(9)}   }
\startdata  
CFg             & $>0.86$            & $>$0.95        & $7.8-8.7$& A      & 23.1 & -21.6     & G & 2.313  & 59 \\
Lynx~2-9691     & 0.70-1.02          & $>$0.77        & $8.3-8.8$& A      & 22.7 & -22.5     & H & 2.888  & 111 \\
MS1512-cB58     & 0.92               & 0.17           & $8.2-8.5$& B      & 23.0 & -22.1     & I & 2.739  & 21 \\
Q0201+113 C6    & $>$0.83            & \nodata        & $7.9-8.7$& C,D    & 23.4 & -22.0     & J & 3.053  & 85 \\
DSF2237+116 C2  & $>$0.65            & \nodata        & $7.6-8.8$& C,E    & 22.3 & -23.3     & J & 3.333  & 276 \\
B0902+343 C6   & $>$0.56             & \nodata        & $7.5-8.9$& C      & 22.9 & -22.5     & J & 3.099  & 53 \\
Westphal CC13   & $>$0.63            & \nodata        & $7.6-8.8$& F      & 23.1 & -22.5     & K & 3.406  & 39 \\
\enddata  
\tablerefs{
(1) Emission line ratio $R_{23}=(F_{4959}+F_{5007}+F_{3727})/(F_{H\beta})$;
(2) Ionization parameter ratio $O_{32}=(F_{5007}+F_{4959})/F_{3727}$;
(3) Oxygen abundance relative to hydrogen;
(4) Reference;
(5) $K_{AB}$ magnitude;
(6) Rest frame blue magnitude on the Vega system for comparison to local galaxies,
computed as described in Column 7;
(7) Reference for photometry; 
(8) Redshift; 
(9) Star formation rates, from either the Balmer recombination line
luminosities for objects presented here, or from [O~III] line
luminosities as tabulated by Teplitz \etal\ (2000a)\\
(A) This work;
(B) Teplitz \etal\ (2000a);
(C) Pettini \etal\ (1998); 
(D) $\left[O~III\right]$ $\lambda$5007 is not observed due to night sky line
contamination; assume $F_{5007}=2.9F_{4959}$;
(E) $\left[O~III\right]$ $\lambda$4959 not observed due to night sky line
contamination; assume $F_{4959}=F_{5007}/2.9$.
(F) Teplitz et. al 2000b;
(G) Computed using the observed $V_{AB}\simeq{V_{555}}=24.14$ from Roche \etal\ (2000), 
assuming the same color as MS1512-cB58 ($V_{AB}-K_{AB}=1.0$) (Ellingson \etal\ 1996) 
to obtain the $K_{AB}$ listed in column 6.  
The distance modulus of 46.4 mag and bandpass term of the K correction, 
-2.5log(1+z)=1.3, 
yields $M_{R~AB}=-22.0$.  Assuming $(B-R_C)_{AB}=0.4$ for an Im type SED (Fukugita \etal\ 1995), 
yields the final restframe B-band estimate of -21.6.
(H) Computed from the measured photographic magnitude of $R_F=23.04$, 
assuming $R_F-R_{AB}\sim-0.1$ and the same $R_{AB}-K_{AB}=0.4$ color as a similar high-redshift object 0000-263 D6
from Pettini \etal\ (1998) yields $K_{AB}=22.7$. 
A distance modulus of 47.0 mag, with the bandpass term of the
K correction -2.5log(1+z)=1.5 mag for $z=2.9$,
yields a rest-frame $M_{V~AB}$ of -22.8.  Adopting $B-V$=0.27 for an Im type SED (Fukugita \etal\
1995) suggests a final rest-frame $M_B$ of -22.5;
(I) Corrected for gravitational magnification (Seitz \etal\ 1998) who give $R=23.9$ and
$K_{AB}=23.0$ after demagnification.  Applying a distance modulus of 46.8
mag, a K correction bandpass term of 1.4 mag, and assuming B-V=0.27 for an Im type 
spectral energy distribution (Fukugita \etal\ 1995)
yields the final rest-frame B magnitude of -22.1; 
(J) Computed assuming that the observed $K_{AB}$ magnitude from Pettini
\etal\ (1996) is approximately the rest frame V magnitude, adopting
B-V=0.27 for an Im type spectral energy distribution (Fukugita \etal\
1995), a K correction of 1.5 mag (1.6 mag for Westphal CC13) , and the appropriate distance modulus;
(K) Computed assuming that the observed $K_{AB}$ magnitude from Teplitz \etal\ 
(2000a) is approximately the rest frame V magnitude,  adopting
B-V=0.27 for an Im type spectral energy distribution (Fukugita \etal\ 1995), a K correction of 1.5 mag, 
and the appropriate distance modulus.
}
\end{deluxetable}

\end{document}